\documentclass[10pt,twocolumn,superscriptaddress,amssymb,amsmath,aps,pra]{revtex4-1}
\usepackage{graphicx}

\begin{document}

\title{Photon mass drag and the momentum of light in a medium}
\date{June 29, 2017}
\author{Mikko Partanen}
\affiliation{Engineered Nanosystems group, School of Science, Aalto University, P.O. Box 11000, 00076 Aalto, Finland}
\author{Teppo H\"ayrynen}
\affiliation{DTU Fotonik, Department of Photonics Engineering, Technical University of
Denmark, \O rsteds Plads, Building 343, DK-2800 Kongens Lyngby, Denmark}
\author{Jani Oksanen}
\affiliation{Engineered Nanosystems group, School of Science, Aalto University, P.O. Box 11000, 00076 Aalto, Finland}
\author{Jukka Tulkki}
\affiliation{Engineered Nanosystems group, School of Science, Aalto University, P.O. Box 11000, 00076 Aalto, Finland}

\begin{abstract}
Conventional theories of electromagnetic waves in a medium assume that
the energy propagating with the light pulse in the medium is entirely carried by
the field. Thus, the possibility that the optical force field of the light pulse
would drive forward an atomic mass density wave (MDW) and the related
kinetic and elastic energies is neglected.
In this work, we present foundations of a covariant theory of light propagation
in a medium by considering a light wave simultaneously with the dynamics
of the medium atoms driven by optoelastic forces
between the induced dipoles and the electromagnetic field. We show that a light pulse having a total
electromagnetic energy $\hbar\omega$
propagating in a nondispersive medium transfers a mass equal to
$\delta m=(n^2-1)\hbar\omega/c^2$, where $n$ is the refractive index.
MDW, which carries this mass, consists of atoms, which are more densely spaced inside the light pulse
as a result of the field-dipole interaction.
We also prove that the transfer of mass with the light pulse, the
photon mass drag effect, gives an essential contribution
to the total momentum of the light pulse, which becomes equal to the
Minkowski momentum $p_\mathrm{M}=n\hbar\omega/c$.
The field's share of the momentum is the Abraham momentum
$p_\mathrm{A}=\hbar\omega/(nc)$, while the difference $p_\mathrm{M}-p_\mathrm{A}$
is carried by MDW. Due to the coupling of the field and matter, only the total
momentum of the light pulse and the transferred mass $\delta m$ can be directly measured.
Thus, our theory gives an unambiguous physical meaning to the Abraham and Minkowski momenta.
We also show that to solve
the centenary Abraham-Minkowski controversy of the momentum of
light in a nondispersive medium in a way that is consistent with Newton's
first law, one must account for the mass transfer effect.
We derive the photon mass drag effect using two independent but complementary covariant models.
In the mass-polariton (MP)
quasiparticle approach, we consider the light pulse as a coupled state
between the photon and matter, isolated from the rest of the medium.
The momentum and the transferred mass of MP follow unambiguously
from the Lorentz invariance and the fundamental conservation laws of nature.
To enable the calculation of the mass and momentum distribution of a light pulse,
we have also generalized the electrodynamics of continuous media
to account for the space- and time-dependent optoelastic dynamics
of the medium driven by the field-dipole forces.
In this optoelastic continuum dynamics (OCD) approach,
we obtain with an appropriate space-time discretization a
numerically accurate solution of the Newtonian continuum dynamics of the medium
when the light pulse is propagating in it. The OCD simulations
of a Gaussian light pulse propagating in a diamond crystal
give the same momentum $p_\mathrm{M}$
and the transferred mass $\delta m$ for the light pulse as the MP quasiparticle approach.
Our simulations also show that, after photon transmission,
some nonequilibrium of the mass distribution is left in the medium. 
Since the elastic forces are included in our simulations
on equal footing with the optical forces, our simulations
also depict how the mass and thermal equilibria are
reestablished by elastic waves.
In the relaxation process, a small amount of photon energy
is dissipated into lattice heat.
We finally discuss a possibility of an optical waveguide setup for experimental
measurement of the transferred mass of the light pulse.
Our main result that a light pulse is inevitably associated with an experimentally measurable mass
is a fundamental change in our understanding of light propagation in a medium.
\end{abstract}

\maketitle

\onecolumngrid
\vspace{-0.2cm}
\twocolumngrid

\section{Introduction}
\vspace{-0.2cm}

\begin{figure}
 \includegraphics[width=0.48\textwidth]{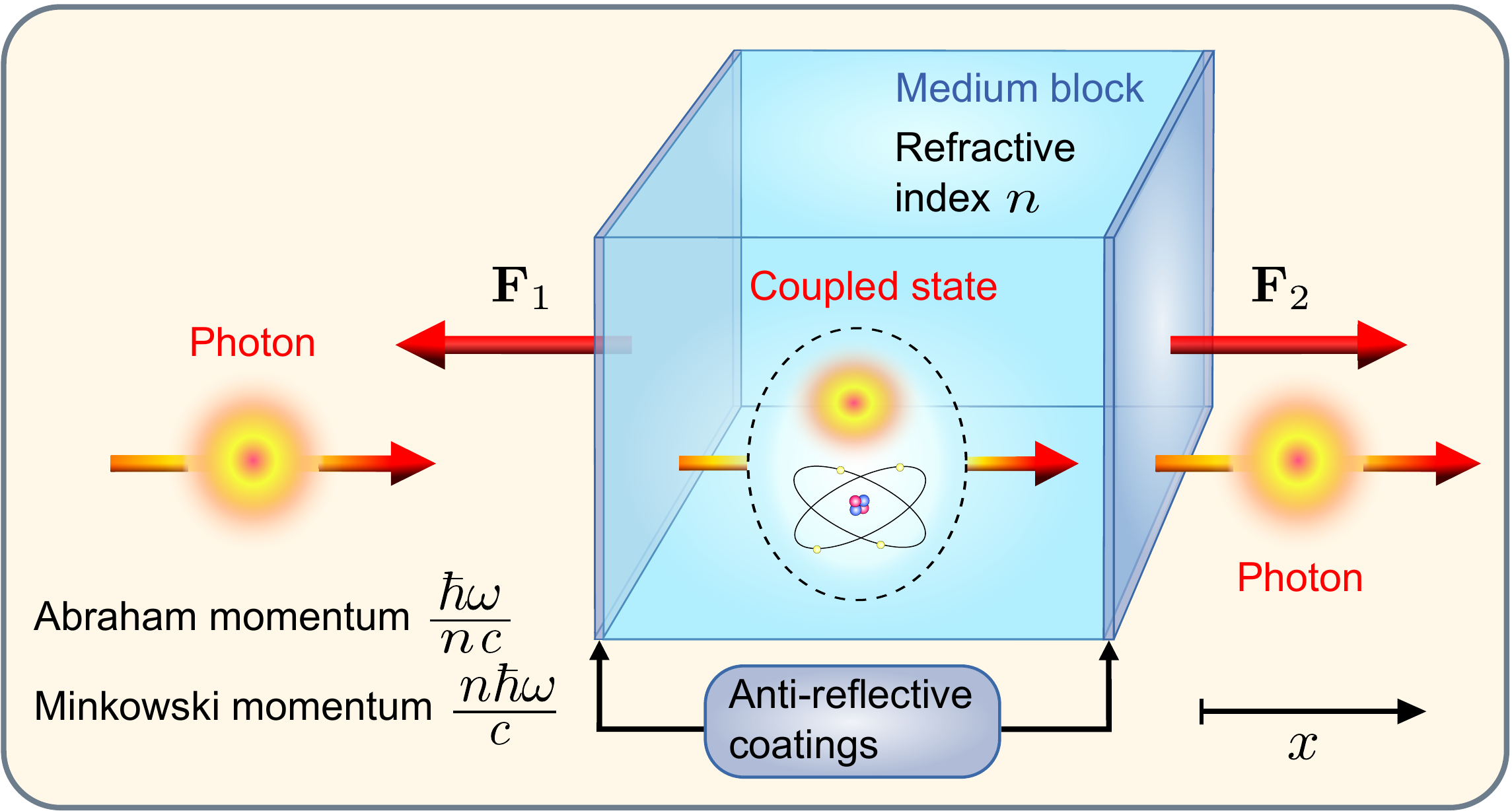}
\caption{\label{fig:illustration}
(Color online) Schematic illustration of a photon propagating
through a medium block with refractive index $n$. Left: the photon is incident
from vacuum. Middle: inside the medium, the photon couples to atoms
forming a quasiparticle,
which continues to propagate inside the block.
Right: at the end of the
block, the photon continues to propagate in vacuum. At the photon entrance
and exit, the medium block experiences recoil forces $\mathbf{F}_1$ and $\mathbf{F}_2$
that depend on the total momentum of light in a medium.
The anti-reflective coatings are included only to simplify the conceptional
understanding of the problem.}
\vspace{-0.2cm}
\end{figure}

Since the introduction of the photon hypothesis by Planck in 1900 \cite{Planck1900},
the relation between energy $E$ and momentum $p$ of a photon propagating in vacuum has been known
to be $E=\hbar\omega=cp$, where $\hbar$ is the reduced Planck constant,
$\omega$ is the angular frequency, and $c$ is the speed of light in vacuum.
This energy-momentum ratio of a photon is one of the experimentally most accurately known
quantities in physics as both the energy and momentum of a photon are accessible to accurate
experimental measurements. In this retrospect, it is astonishing that the momentum of
a light wave propagating in a medium has remained a subject of an extensive
controversy until now.
This controversy is known as the Abraham-Minkowski dilemma
\cite{Leonhardt2006,Cho2010,Pfeifer2007,Barnett2010b,Barnett2010a,Kemp2011,Leonhardt2014,Milonni2010,Pfeifer2009,Brevik2017,Ramos2015,Testa2013,Crenshaw2013}.
The rivaling momenta of light in a medium are $p_\mathrm{A}=\hbar\omega/(nc)$ (Abraham) \cite{Abraham1909,Abraham1910}
and $p_\mathrm{M}=n\hbar\omega/c$ (Minkowski) \cite{Minkowski1908},
where $n$ is the refractive index.
In this work, we will show that the Abraham-Minkowski dilemma is just one of
the enigmas following essentially from the breakdown of covariance condition
in the existing theories of light  propagation in a medium.
The problem of light propagation in a medium is schematically illustrated in Fig.~\ref{fig:illustration}.

The origin of the Abraham-Minkowski controversy may be traced back to difficulties in
generalizing the Planck's photon hypothesis into a quantum
electrodynamical description of light wave propagation in a medium.
Extensive research effort has continued until now also on
experimental determination of the photon momentum and
several different experimental setups have been introduced
\cite{Campbell2005,Sapiro2009,Jones1954,Jones1978,Walker1975,She2008,Zhang2015,Astrath2014,Ashkin1973,Casner2001,Brevik1979}.
However, accurate experimental determination of the momentum
of light in a medium has proven to be surprisingly difficult.
On the theoretical side, neither the Abraham nor the Minkowski momentum
has been proven to be fully consistent with the energy and momentum conservation
laws, Lorentz invariance, and available experimental data.
To explain the ambiguities, it has also been suggested that both forms of
momenta are correct but simply represent different aspects of the photon momentum
\cite{Barnett2010b,Barnett2010a}.
In some other works, the Abraham-Minkowski controversy has
been claimed to be resolved by arguing that division of the total
energy-momentum tensor into electromagnetic and material components
would be arbitrary \cite{Pfeifer2007,Penfield1967,deGroot1972}.

It is well known in the electrodynamics
of continuous media that, when a light pulse propagates in a medium,
the medium atoms are a subject of field-dipole forces \cite{Landau1984}.
However, the dynamics of a light pulse, a coupled state of the field and matter, driven by
the field-dipole forces has been a subject of very few detailed studies.
In this work, we elaborate how these driving forces give rise to a
mass density wave (MDW) in the medium when a light pulse is
propagating in it. MDW is studied using two independent approaches.
The first approach is the mass-polariton (MP) quasiparticle picture.
In this picture, the coupled state of the field and matter is considered
isolated from the rest of the medium and thus a subject of the covariance
principle and the general conservation laws of nature. In the second
approach, we apply the electrodynamics of continuous media and
continuum mechanics to compute the dynamics of the medium
when a light pulse is propagating in it.
In this optoelastic continuum dynamics (OCD) approach, the total force on a medium element consists
of the field-dipole force and the elastic force resulting from
the density variations in the medium.
In brief, we will show that accounting for MDW resulting from the coupled dynamics
of the field and matter allows formulating a fully consistent covariant theory
of light in a medium.
The covariant theory also gives a unique resolution to
the Abraham-Minkowski controversy.

This paper is organized as follows:
Section \ref{sec:experiments} briefly summarizes some of the most conclusive experiments related to the
determination of the momentum of light in a medium. Section \ref{sec:covarianceproblem}
presents an introduction to the covariance problem in the conventional theories
of light in a medium. In Sec.~\ref{sec:mptheory}, we present the MP
quasiparticle model using the Lorentz transformation and general conservation laws.
In Sec.~\ref{sec:continuum},
we derive the OCD model by coupling the electrodynamics of continuous
media with the continuum mechanics to describe the propagation of the coupled
state of the field and matter in a medium.
Numerical OCD simulations of MDW are presented in Sec.~\ref{sec:simulations}.
The results of the MP quasiparticle picture and the
OCD method are compared in Sec.~\ref{sec:comparison}.
In Sec.~\ref{sec:discussion}, we discuss the significance of our results
to the theory of light propagation in a medium and
compare our theory to selected earlier theoretical works.
We also discuss the possibility of direct experimental
verification of MDW and the covariant state of light in a medium.
Finally, conclusions are drawn in Sec.~\ref{sec:conclusions}.

\section{\label{sec:experiments}Brief summary of experiments}

\begin{figure}
\includegraphics[width=0.48\textwidth]{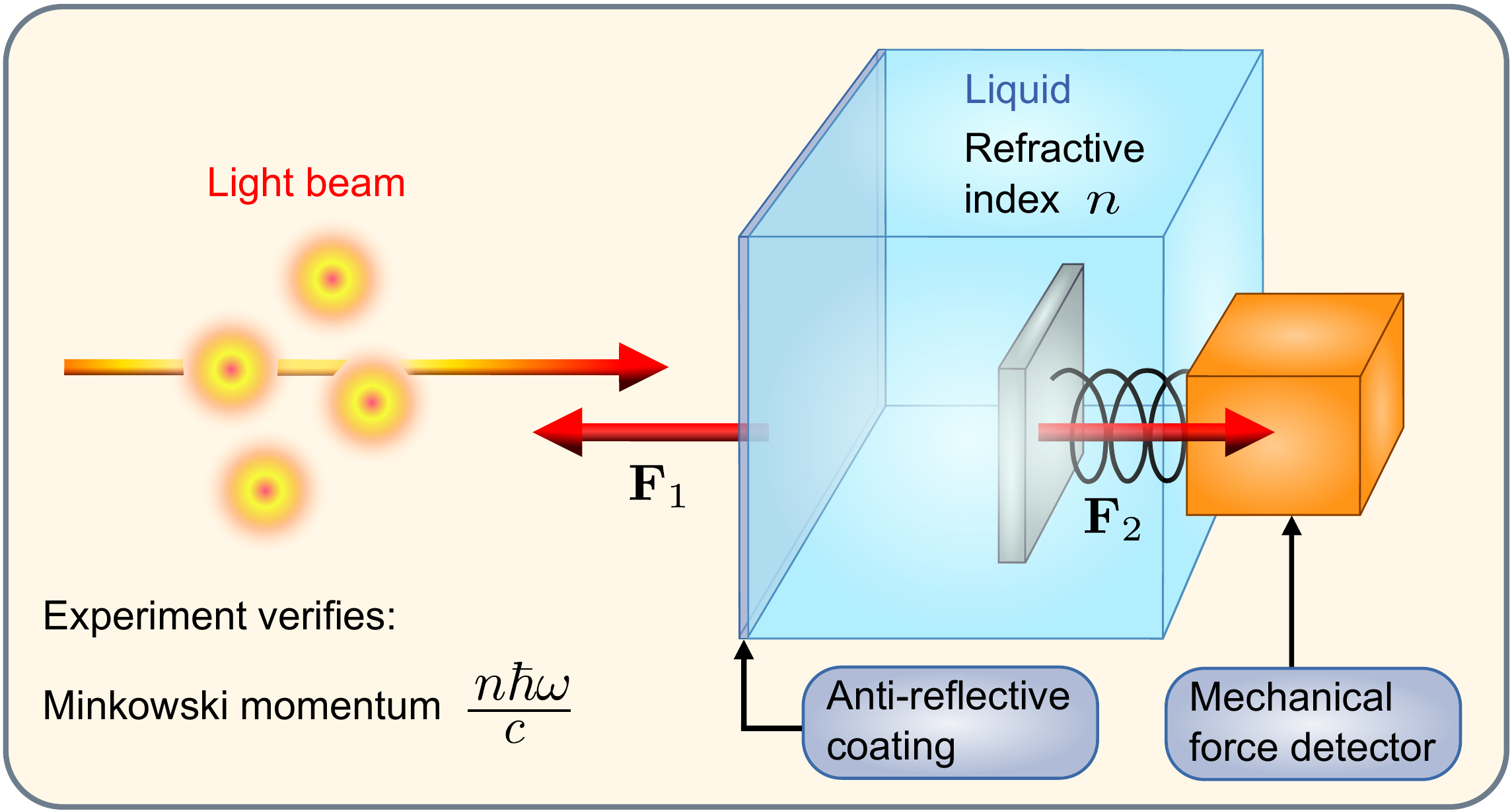}
\caption{\label{fig:reflectorscheme}
(Color online) Schematic illustration of an experimental setup for
the measurement of the electromagnetic forces due to a light beam. Light enters from vacuum
to a liquid container with anti-reflective coating. Inside the liquid
having refractive index $n$, light is fully reflected
from a mirror attached to a detector that measures the resulting force.
The force is found to be proportional to the refractive index in
accordance with the Minkowski momentum.}
\vspace{-0.2cm}
\end{figure}

To find the correct form for the light momentum, numerous experiments have been carried out.
The most conclusive set of experiments were started in 1954 by
Jones and Richards \cite{Jones1954} who studied the pressure exerted
by light on a reflector immersed in a liquid with refractive index $n$.
They showed with 1\% precision that the pressure
on a reflector immersed in a liquid is $n$ times
the pressure exerted on the reflector by the same light in free space.
The experiment was repeated in 1978 by Jones and Leslie \cite{Jones1978} with
0.05\% precision. A simplified, but principally identical setup is
illustrated in Fig.~\ref{fig:reflectorscheme}. In the setup,
the photon is incident from vacuum to dielectric liquid.
The first interface of the liquid container has anti-reflective coating
to avoid reflections and the second interface is a perfect reflector
attached to a mechanical force detector.
If we assume that the force on the reflector
is completely determined by the sum of impulses $\Delta\mathbf{p}_i$ of each photon
in time $\Delta t$
as $\mathbf{F}_2=\sum_i\Delta\mathbf{p}_i/\Delta t$,
then the experiment supports
the Minkowski formula $p=n\hbar\omega/c$
\cite{Pfeifer2007,Barnett2010b,Barnett2010a}.

There exist also other experiments where the Minkowski form
of the momentum has been verified including the investigations of the recoil
of atoms in a dilute Bose-Einstein condensate gas \cite{Campbell2005,Sapiro2009} and
the conventional absorption related photon drag effect, which
generates currents and electric fields in semiconductors by the transfer
of light momentum to the free charge carriers \cite{Gibson1970,Loudon2005,Mansuripur2005}.
The Minkowski form of momentum also correctly
explains the refractive index dependence
of the Doppler shift in dilute gases \cite{Bradshaw2010,Milonni2005}.
There does not seem to exist any direct, quantitative, and reproducible measurements
reporting the Abraham momentum for a propagating field
inside the medium. All measurements
that have been so far argued to support the Abraham
momentum \cite{Walker1975,She2008,Zhang2015}
measure the movement of the medium
or its surface from outside. Therefore, those measurements
do not measure forces directly related to the
total momentum of a light wave inside the medium. Instead, possible
reflections, surface forces,
and the MDW effect predicted in this work 
should all be taken into account in their correct
interpretation.
The same interpretation problem exists also in the
surface deformation measurements
\cite{Astrath2014,Ashkin1973,Casner2001}
that have been claimed to support
the Minkowski momentum.

\section{\label{sec:covarianceproblem}Covariance problem}

We first consider how the so-called Einstein's box thought experiment
\cite{Einstein1906,Balazs1953}
is traditionally applied to determine the photon momentum
inside the medium.
The starting point is the generalized Newton's first law \cite{Barnett2010b},
also known as the constant center-of-energy velocity (CEV) for
an isolated system like the photon plus the medium in Fig.~\ref{fig:illustration}.
The constant CEV 
for a system of a photon with energy $\hbar\omega$ and velocity $c$ initially in vacuum
and a medium block with mass $M$ and energy $Mc^2$ initially at rest
is conventionally argued to obey the equation
\begin{equation}
 V_\mathrm{CEV}=\frac{\sum_iE_iv_i}{\sum_iE_i}=\frac{\hbar\omega c}{\hbar\omega+Mc^2}=\frac{\hbar\omega v+Mc^2V}{\hbar\omega+Mc^2},
 \label{eq:uniformmotion}
\end{equation}
where the first and second forms are written
for the cases before and after the photon has entered the medium.
It is assumed that inside the medium the initial photon
energy $\hbar\omega$ propagates with velocity $v=c/n$ which
results in the medium block obtaining a velocity $V$ to be determined from Eq.~\eqref{eq:uniformmotion}.
From Eq.~\eqref{eq:uniformmotion} we obtain
$\hbar\omega/c=\hbar\omega/(nc)+MV$
suggesting that the initial photon momentum $\hbar\omega/c$ in vacuum
is split to the Abraham momentum of a photon in a medium equal to $\hbar\omega/(nc)$
and to the medium block momentum equal to $MV$.
One might then conclude that
the Abraham momentum would be the correct photon momentum in a medium.

However, the above result leads to a striking contradiction
with the covariance principle,
which is a fundamental prerequisite of the special theory of relativity.
The reader can easily verify that a photon with energy $E=\hbar\omega$
and momentum $p=\hbar\omega/(nc)$
does not fulfill the covariance condition $E^2-(pc)^2=(m_\mathrm{ph}c^2)^2$
if the rest mass $m_\mathrm{ph}$ of a photon is set to zero.
The same contradiction exists also with the conventional
definition of the Minkowski momentum.
Since earlier formulations of the theory assuming zero rest mass
have failed to lead to a covariant description of the light wave, it is natural to consider
a possibility that the light wave is actually a coupled state of
the field and matter with a small but finite rest mass.
As we will show below,
this rest mass will originate from the atomic mass density
wave that is driven forward by the field-dipole forces
associated with the light pulse.

\section{\label{sec:mptheory}Mass-polariton theory}

In this section, we present an unambiguous MP quasiparticle model of light in
a medium. The MP model fully satisfies
the fundamental conservation laws of nature
and the covariance condition, which follows from all laws of physics
being the same for all inertial observers.
We assume that the medium is nondispersive for the band of wavelengths discussed,
typically transparent solid or liquid.
Generalizing Feynman's description of light propagating in solids
\cite{Feynman1964}, we show
that the light quantum must form a coupled state with the atoms
in the medium. This MP quasiparticle is shown to have a rest mass
that propagates through the medium at speed $v=c/n$.
Hence we use the polariton concept in a meaning that
is fundamentally different from its conventional use in the context of
the phonon-polariton and the exciton-polariton quasiparticles.
In these latter cases, a photon propagating in a medium is in resonance
or in close resonance with an internal excited state of the medium.

Given the energy of the incoming photon in vacuum $\hbar\omega$,
we start by calculating which part of this energy
is transmitted to the coupled state of the field and matter, MP,
that continues to propagate in the medium.
Since it is possible that, at the left interface of the medium in Fig.~\ref{fig:illustration},
the momentum
of the photon may change from its vacuum value, the thin interface layer
of the medium has to take the recoil momentum and the related kinetic recoil energy
to balance the momentum and energy conservation laws.
In this work, with \emph{recoil} energies and momenta we consistently mean 
the energies and momenta taken by the thin interface layers when
the photon enters or escapes the medium.
We can easily show that the recoil energy is negligibly small
in comparison with $\hbar\omega$.
An estimate of the recoil energy
is given by $E_\mathrm{a}=P_\mathrm{a}^2/(2M_\mathrm{a})$
where $M_\mathrm{a}$ is the total mass of the recoiling surface atoms and
$P_\mathrm{a}$ is the recoil momentum.
Since the momentum of MP is unknown for the moment,
we cannot know the exact value of $P_\mathrm{a}$.
However, we can certainly use the momentum of the 
incoming photon as an \emph{order-of-magnitude} estimate for $P_\mathrm{a}$.
Setting $P_\mathrm{a}$ equal to $\hbar\omega/c$
results in $E_\mathrm{a}/\hbar\omega=\hbar\omega/(2M_\mathrm{a}c^2)$,
which is extremely small.
Thus, MP gains the whole field energy $\hbar\omega$ of the incoming photon.
In the OCD simulations in Sec.~\ref{sec:simulations},
we also compute the recoil energy to check the accuracy of this approximation.
Note that the recoil momentum of the body, which is so far unknown,
\emph{is not needed in the following analysis}.

It is impossible to fulfill the covariance condition within a zero rest
mass assumption if the momentum of MP is not equal to 
the vacuum value of the photon momentum $\hbar\omega/c$.
We next study a possibility of a covariant state of a light pulse in
a medium having a momentum $p_\text{\tiny MP}$, which \emph{a priori} is not equal to $\hbar\omega/c$.
Accordingly, we assume that a mass $\delta m$, which is
a part of the initial medium block mass $M$, is associated with MP.
We will determine the value of $\delta m$ by requiring that 
MP is described by a covariant state that enables the transfer of energy trough the 
medium at speed $v=c/n$. We will show that by determining the value of $\delta m$, we also 
determine the value of the momentum of MP.
Therefore, the ratio $\delta mc^2/\hbar\omega$
and the energy-momentum ratio $E_\text{\tiny MP}/p_\text{\tiny MP}$ of MP are \emph{internal properties}
of the light pulse.
\emph{Interface forces} in Fig.~\ref{fig:illustration} are needed to balance
the \emph{momentum of the covariant state} of MP.
The covariance principle gives unambiguous values for the momentum
and the rest mass of the mass polariton and also the interface forces
$\mathbf{F}_1$ and $\mathbf{F}_2$ in Fig.~\ref{fig:illustration}.
However, the microscopic distribution of the recoil momenta near the
interfaces of the medium block can only be calculated by using
a microscopic theory.

\begin{figure}
\includegraphics[width=0.48\textwidth]{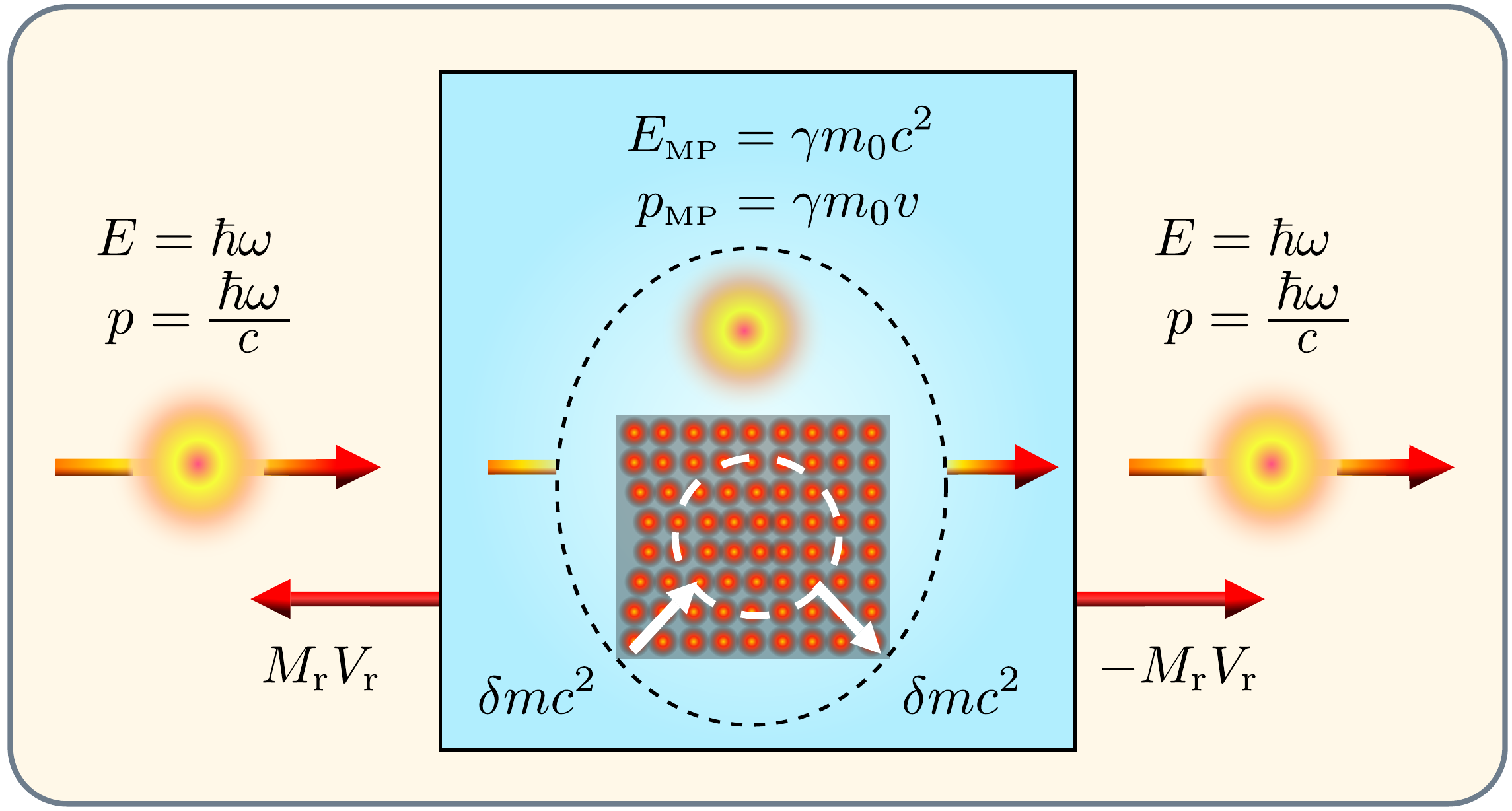}
\caption{\label{fig:diagram}
(Color online) Illustration of the energy and momentum
transfer between the photon and the medium block in the processes where
MP is created and destroyed.}
\vspace{-0.2cm}
\end{figure}

In the actual medium block, the mass $\delta m$
is associated to MDW,
which describes the displacement of atoms
inside the light pulse as illustrated in Fig.~\ref{fig:diagram}.
The exact space distribution of $\delta m$ depends on
the shape of the light pulse and the detailed material properties and it can be calculated
numerically using OCD discussed in Sec.~\ref{sec:continuum}. 
However, the total mass $\delta m$ transferred can be determined by using the
Lorentz transformation and the Doppler shift of the photon energy as shown below. We start by
considering the MP energy in the laboratory frame.

\emph{Laboratory frame (L frame)}.  The total energy of MP
can be split into two contributions:
(1) the energy of the field including the potential energy of induced dipoles and
the kinetic energy of MDW atoms
equal to $\hbar\omega$ and
(2) the mass energy of $\delta m$
transferred with MDW equal to $\delta mc^2$.
In the total energy of MP in the L frame, all other contributions,
but the mass energy $\delta mc^2$,
have their origin in the field energy of the incoming photon.
In any inertial frame, all this field energy can be exploited,
for example, in the resonance excitation of the medium atoms.

It is tempting to think that the 
kinetic energy of MDW is given by $(\gamma-1)\delta mc^2$,
where $\gamma=1/\sqrt{1-v^2/c^2}$ is the Lorentz factor, i.e., the mass $\delta m$
would be moving at the speed $v=c/n$ through the medium.
However, this is not the case.
Since $\delta m$ is carried by MDW, a wave packet made of \emph{increased} atomic density,
the total mass of atoms in the wave packet is vastly larger
than $\delta m$.
In analogy with the discussion of the recoil energy above,
it is straightforward to show that
the kinetic energy of atoms in MDW remains negligibly small due to their 
large rest energy in comparison with $\hbar\omega$.
The understanding 
of  this  seemingly  non-intuitive  result  can
be  facilitated  by  a schematic model in which the wave of atoms bound 
together by elastic forces is driven forward by
the electromagnetic field at the speed $v=c/n$.
Thus, MDW should not be confused with the sound
wave, which is driven forward by elastic forces.
In the OCD method
in Sec.~\ref{sec:continuum}, we
also account for the very small kinetic energies
of atoms in MDW and show that
this picture is in full accordance with
the semiclassical theory of the electromagnetic field
in a medium and the continuum mechanics.

\emph{MP rest frame (R frame)}.
When the total L-frame energy of MP is transformed to an inertial
frame moving with the velocity $v'$ with respect to L frame,
we obtain the total energy of MP using the Lorentz transformation
\begin{equation}
 E_\text{\tiny MP}\longrightarrow\gamma'(E_\text{\tiny MP}-v'p_\text{\tiny MP})=\gamma'(\hbar\omega-v'p_\text{\tiny MP})+\gamma'\delta mc^2,
 \label{eq:LorentzE}
\end{equation}
\vspace{-0.7cm}
\begin{equation}
 p_\text{\tiny MP}\longrightarrow\gamma'\Big(p_\text{\tiny MP}-\frac{v'E_\text{\tiny MP}}{c^2}\Big)=\gamma'\Big(p_\text{\tiny MP}-\frac{v'\hbar\omega}{c^2}-v'\delta m\Big),
 \label{eq:Lorentzp}
\end{equation}
where $E\text{\tiny MP}=\hbar\omega+\delta mc^2$ is the total energy
of MP in L frame, $p_\text{\tiny MP}$ is the so far unknown momentum of MP
in L frame, and $\gamma'=1/\sqrt{1-v'^2/c^2}$.
In Eq.~\eqref{eq:LorentzE}, the last term on the right
presents the transformed rest energy of MP,
while the first term $\hbar\omega'=\gamma'(\hbar\omega-v'p_\text{\tiny MP})$ has its origin entirely
in the field energy and is equal to the Doppler-shifted energy
of a photon in a medium \cite{ChenJi2011}.
The field energy accessible to the excitation of the medium atoms disappears in the
reference frame moving with the velocity of light in the medium.
Therefore, $\hbar\omega'\longrightarrow 0$ in R frame where $v'=c/n$ and
we obtain
\begin{equation}
p_\text{\tiny MP}=\frac{n\hbar\omega}{c}.
\label{eq:momentum}
\end{equation}
Thus, the Minkowski momentum of MP follows directly from the
Doppler principle \cite{Barnett2010b,Milonni2005},
which, however, must be used as a part of the Lorentz transformation
in Eq.~\eqref{eq:LorentzE} in order to enable the determination of the
transferred mass $\delta m$ of MP.

Since R frame moves with MP, the total momentum of MP is zero
in R frame. Therefore, inserting momentum $p_\text{\tiny MP}$
from Eq.~\eqref{eq:momentum} to Eq.~\eqref{eq:Lorentzp} and
setting $v'=c/n$, we obtain
\begin{equation}
 \delta m=(n^2-1)\hbar\omega/c^2.
 \label{eq:mdwmass1}
\end{equation}
According to the special theory of relativity, we can write the total
energy of MP in its rest frame as $m_0c^2$, where $m_0$ is the
rest mass of the structural system of MP. Therefore, inserting
$p_\text{\tiny MP}$ and $\delta m$
in Eqs.~\eqref{eq:momentum} and \eqref{eq:mdwmass1} to Eq.~\eqref{eq:LorentzE}, we obtain in
R frame
\begin{equation}
 m_0 =n\sqrt{n^2-1}\,\hbar\omega/c^2.
 \label{eq:restmass}
\end{equation}

For the total energy and momentum of MP, one obtains in L frame
\begin{align}
 E_\text{\tiny MP} &=\gamma m_0c^2=n^2\hbar\omega\nonumber,\\
 p_\text{\tiny MP} &=\gamma m_0v=\frac{n\hbar\omega}{c}.
 \label{eq:results}
\end{align}
The Minkowski form of the MP momentum
and the transferred mass follow directly from
the Lorentz transformation, Doppler shift, and the fundamental
conservation laws of nature.
This is in contrast with earlier explanations
of the Minkowski momentum,
where the possibility of a nonzero
transferred mass carried by MDW
has been overlooked.
For the experimental verification of the covariant state 
of light in a medium, one has to measure both the momentum and
the transferred mass of MP.

The energy and momentum in Eq.~\eqref{eq:results} and the rest mass in Eq.~\eqref{eq:restmass}
fulfill the covariance condition $E_\text{\tiny MP}^2-(p_\text{\tiny MP}c)^2=(m_0c^2)^2$,
thus resolving the covariance problem discussed in Sec.~\ref{sec:covarianceproblem}. 
Note that although knowing $\delta m$ is enough to understand the mass transfer
associated with MP, $m_0$ is useful
for transparent understanding of the 
covariant state of light in a medium.
For an additional discussion on
the relation between $\delta m$ and $m_0$ in
the Lorentz transformation, see Appendix \ref{apx:DopplerLorentz}.

The covariant energy-momentum ratio $E/p=c^2/v$ also
allows splitting the total MP momentum in Eq.~\eqref{eq:results}
into parts corresponding to the electromagnetic energy $\hbar\omega$
and the MDW energy $\delta mc^2$. As a result, the field
and MDW momenta are given in L frame by
\begin{align}
 p_\mathrm{field} &=\frac{\hbar\omega}{nc},\nonumber\\
 p_\text{\tiny MDW} &=\Big(n-\frac{1}{n}\Big)\frac{\hbar\omega}{c}.
 \label{eq:momentumsplitting}
\end{align}
Therefore, the field's share of the total MP momentum
is of the Abraham form while the MDW's share is given
by the difference of the Minkowski and Abraham momenta.
However, due to the coupling, only the total momentum
of MP and the transferred mass are directly measurable.

Our results in Eqs.~\eqref{eq:restmass} and \eqref{eq:results} also show that
the rest mass $m_0$ has not been taken
properly into account in the Einstein's box thought experiment
discussed above. Accounting for the rest mass of MP
allows writing the constant CEV law
in Eq.~\eqref{eq:uniformmotion} before and after
the photon has entered the medium as
\begin{equation}
 V_\mathrm{CEV}=\frac{\sum_iE_iv_i}{\sum_iE_i}=\frac{\hbar\omega c}{\hbar\omega+Mc^2}=\frac{\gamma m_0c^2v+M_\mathrm{r}c^2V_\mathrm{r}}{\gamma m_0c^2+M_\mathrm{r}c^2},
 \label{eq:uniformmotion2}
\end{equation}
where $M_\mathrm{r}=M-\delta m$ and $V_\mathrm{r}=(1-n)\hbar\omega/(M_\mathrm{r}c)$.
The equality of the denominators is nothing but
the conservation of energy,
and the equality of the numerators divided by $c^2$ corresponds to the
momentum conservation.
Equation \eqref{eq:uniformmotion2} directly shows that MP
with the Minkowski momentum obeys the constant CEV motion and explains
why earlier derivations of the Minkowski momentum assuming
zero rest mass for the light pulse lead to violation
of the constant CEV motion \cite{Barnett2010b,Barnett2010a}.

\section{\label{sec:continuum}Optoelastic continuum dynamics}

Above, we have derived the MP quasiparticle model using the fundamental conservation
laws and the Lorentz transformation. To give a physical meaning for the
transferred mass $\delta m$ and the momentum of MDW, we next
present the complementary OCD model based on the electrodynamics of continuous
media and the continuum mechanics. The OCD model enables the
calculation of the mass and momentum distribution of a light pulse
as a function of space and time.
In contrast to the MP model above,
here we also account for the very small kinetic energies of atoms
in MDW and the recoil energies at the interfaces.
It is surprising that, although the classical theories of electrodynamics
and continuum mechanics are well known,
they have not been combined to compute the propagation of an optical light pulse
and the associated MDW in a medium.
Note that, in the OCD model, there is no need to separately include the Lorentz transformation,
conservation laws, or Doppler shift since they are inherently accounted for
by the Maxwell's equations and the elasticity theory.
In calculating the optoelastic force field, we assume that the damping
of the electromagnetic field due to the transfer of field energy to the kinetic and elastic
energies of the medium is negligible. We check the accuracy of this approximation
at the end of Sec.~\ref{sec:simulations}.

\subsection{\label{sec:electrodynamics}Energy and momentum in the electrodynamics of continuous media}

\subsubsection{Mass-polariton momentum}

We derive the MP momentum
by using the standard theory of the electrodynamics of continuous media.
In previous literature, there has been extensive discussion on the appropriate form of
the force density acting on the medium under the influence of time-dependent
electromagnetic field \cite{Milonni2010}.
We apply the form of the optical force density that is most widely used
in previous literature \cite{Landau1984,Milonni2010}.
It is given for a dielectric and magnetic fluid by \cite{Landau1984}
\begin{align}
 &\mathbf{f}_\mathrm{opt}(\mathbf{r},t) \nonumber\\
&=-\nabla P
+\frac{1}{2}\nabla\Big[\rho_\mathrm{a}\frac{\partial\varepsilon}{\partial\rho_\mathrm{a}}\mathbf{E}^2\Big]
+\frac{1}{2}\nabla\Big[\rho_\mathrm{a}\frac{\partial\mu}{\partial\rho_\mathrm{a}}\mathbf{H}^2\Big]\nonumber\\
&\hspace{0.5cm}-\frac{1}{2}\mathbf{E}^2\nabla\varepsilon
-\frac{1}{2}\mathbf{H}^2\nabla\mu
+\frac{n^2-1}{c^2}\frac{\partial}{\partial t}\mathbf{S}.
 \label{eq:opticalforcedensity0}
\end{align}
Here $\mathbf{E}$ and $\mathbf{H}$ are the electric and magnetic
field strengths, $\mathbf{S}=\mathbf{E}\times\mathbf{H}$ is the Poynting vector,
$P$ is the field-dependent pressure in the medium,
$\rho_\mathrm{a}$ is the atomic mass density of the medium, and
$\varepsilon$ and $\mu$ are the permittivity and permeability of the medium.
The relative permittivity and permeability of the medium are given by
$\varepsilon_\mathrm{r}=\varepsilon/\varepsilon_0$
and $\mu_\mathrm{r}=\mu/\mu_0$, where $\varepsilon_0$ and $\mu_0$
are the permittivity and permeability of vacuum. They are related
to the refractive index of the medium as $\varepsilon_\mathrm{r}\mu_\mathrm{r}=n^2$.

Assuming a dielectric medium with $\mu_\mathrm{r}=1$
in mechanical equilibrium, it has been previously reasoned
that the first two terms
on the right in Eq.~\eqref{eq:opticalforcedensity0}
cancel each other \cite{Milonni2010,Pitaevskii2006}.
In this case, one obtains \cite{Milonni2010}
\begin{equation}
 \mathbf{f}_\mathrm{opt}(\mathbf{r},t) =-\frac{\varepsilon_0}{2}\mathbf{E}^2\nabla n^2+\frac{n^2-1}{c^2}\frac{\partial}{\partial t}\mathbf{S}.
 \label{eq:opticalforcedensity}
\end{equation}
The second term in Eq.~\eqref{eq:opticalforcedensity}
is known as the Abraham force density \cite{Milonni2010,Brevik1979,Landau1984} and the total force density
in Eq.~\eqref{eq:opticalforcedensity} corresponds to the total force density experienced
by the medium in the conventional Abraham model \cite{Milonni2010,Brevik1979}.
We use this result as given
and describe the relaxation of the extremely small mechanical
nonequilibrium resulting from the optical force density in Eq.~\eqref{eq:opticalforcedensity}
by elastic forces as described in Sec.~\ref{sec:elasticity}.

The force is by definition the time derivative of the momentum.
Therefore, for an electromagnetic pulse propagating in a medium, the
momentum density carried by the induced dipoles becomes
$\mathbf{g}(\mathbf{r},t) = \int_{-\infty}^t\mathbf{f}_\mathrm{opt}(\mathbf{r},t')dt'$ which,
by substituting the force density in Eq.~\eqref{eq:opticalforcedensity}, results in
\begin{equation}
 \mathbf{g}(\mathbf{r},t) =-\frac{\varepsilon_0}{2}\int_{-\infty}^t \mathbf{E}^2\nabla n^2 dt'+\frac{n^2-1}{c^2}\mathbf{S}.
 \label{eq:dipolemomentumdensity}
\end{equation}
The total momentum obtained by the induced dipoles is split into two parts.
The first part has its origin in the change of the refractive index at the interfaces.
Thus it represents the momentum gained by the atoms in the vicinity
of the surfaces. It will be shown in Sec.~\ref{sec:simulations} that this
part of the atomic momentum is directed opposite to the wave vector of the light pulse
on the first interface and parallel to the wave vector on the second interface,
respectively. Thus it corresponds to the recoil forces in Fig.~\ref{fig:illustration}.
The second term corresponds to the momentum of atoms in MDW.
The total momentum carried by MDW is given by a volume integral
of the second term in Eq.~\eqref{eq:dipolemomentumdensity}
over the light pulse.

When we consider a light pulse propagating inside a medium,
the first term of the momentum density in Eq.~\eqref{eq:dipolemomentumdensity}
becomes zero as the refractive index inside the medium is assumed
to be constant with respect to position.
We will consider the first term and the interface effects later in Sec.~\ref{sec:elasticity}.
Therefore, the total momentum carried by MDW
becomes
\begin{equation}
 \mathbf{p}_\text{\tiny MDW} =\int\frac{n^2-1}{c^2}\mathbf{S}d^3r.
 \label{eq:dipolemomentum}
\end{equation}

The total propagating MP momentum is the sum
$\mathbf{p}_\text{\tiny MP}=\mathbf{p}_\text{field}+\mathbf{p}_\text{\tiny MDW}$, where
$\mathbf{p}_\text{field}$ is the momentum carried by
the electromagnetic field.
The momentum density of the field corresponding to the force density
experienced by the medium in Eq.~\eqref{eq:opticalforcedensity},
is known to be of the Abraham form, given by \cite{Landau1984}
\begin{equation}
 \mathbf{p}_\text{field} =\int\frac{1}{c^2}\mathbf{S}d^3r.
 \label{eq:fieldmomentum}
\end{equation}
Thus, the total MP momentum that
is a sum of the MDW and field associated momenta in
Eqs.~\eqref{eq:dipolemomentum} and \eqref{eq:fieldmomentum}
is written as
\begin{equation}
 \mathbf{p}_\text{\tiny MP}=\mathbf{p}_\text{field}+\mathbf{p}_\text{\tiny MDW}=\int\frac{n^2}{c^2}\mathbf{S}d^3r,
 \label{eq:polaritonmomentum}
\end{equation}
which is clearly of the Minkowski form \cite{Barnett2010b}
in agreement with the MP model in Sec.~\ref{sec:mptheory}.

Equation \eqref{eq:polaritonmomentum} states that, when the
Abraham momentum density of the field is added to
the momentum of the medium
calculated from the Abraham force density,
one obtains the Minkowski momentum density.
This result has also been reported earlier \cite{Milonni2005}.
However, in earlier works, the Abraham force density
has not been used to calculate the movement of atoms
and the related transfer of atomic mass with MDW driven by the light pulse.
Thus, the whole of the physical picture of the coupled dynamics of the field
and matter has remained uncovered until now.

\subsubsection{Mass-polariton energy}

Next we write the MP energy and its MDW and field contributions
in terms of the electric and magnetic fields.
Using the covariant energy-momentum ratio
$E/p=c^2/v$ and the relation
$\mathbf{S}=u\mathbf{v}$, where $\mathbf{v}$ is the velocity vector of MP
and $u=\frac{1}{2}(\varepsilon\mathbf{E}^2+\mu\mathbf{H}^2)$
is the energy density of the field
\cite{Jackson1999}, we have
\begin{align}
 E_\text{\tiny MP} &=\int\frac{n^2}{2}(\varepsilon\mathbf{E}^2+\mu\mathbf{H}^2)d^3r,
 \label{eq:polaritonenergy}\\
 E_\text{\tiny MDW} &=\int\frac{n^2-1}{2}(\varepsilon\mathbf{E}^2+\mu\mathbf{H}^2)d^3r,
 \label{eq:dipoleenergy}\\
 E_\text{field} &=\int\frac{1}{2}(\varepsilon\mathbf{E}^2+\mu\mathbf{H}^2)d^3r.
 \label{eq:fieldenergy}
\end{align}
In the coherent field picture, the harmonic field components have to be added
before the calculation of the total energy.
For the application of the above relations for a Gaussian pulse,
see Sec.~\ref{sec:simulations}.

At this stage, one may feel that 
the MDW energy and momentum still
remain as abstract quantities.
However, they will be shown to have a very concrete meaning
in the next section, where we
obtain the MDW energy and momentum by using
Newton's equation of motion for the total mass
density of the medium. This picture also allows numerical
simulations of MDW presented in Sec.~\ref{sec:simulations}.
In Appendix \ref{apx:tensors}, the results of Eqs.~\eqref{eq:dipolemomentum}--\eqref{eq:fieldenergy}
are presented using the well-known energy-momentum
tensor (EMT) formalism.

\subsection{\label{sec:elasticity}Dynamics of the medium in continuum mechanics}

\subsubsection{Newton's equation of motion}

The medium experiences the optical force density $\mathbf{f}_\mathrm{opt}(\mathbf{r},t)$,
which consists of the interaction between the induced dipoles and the
electromagnetic field and effectively also accounts for the interaction between the induced dipoles.
This force can be calculated by using Eq.~\eqref{eq:opticalforcedensity}.
Below, we will show using Newtonian formulation of the continuum mechanics
that this optical force gives rise to MDW and the associated recoil
effect and thus perturbs the mass density of the medium
from its equilibrium value $\rho_0$. The perturbed atomic mass density then becomes
$\rho_\mathrm{a}(\mathbf{r},t)=\rho_0+\rho_\mathrm{rec}(\mathbf{r},t)+\rho_\text{\tiny MDW}(\mathbf{r},t)$,
where $\rho_\mathrm{rec}(\mathbf{r},t)$
is the mass density perturbation due to the recoil effect and $\rho_\text{\tiny MDW}(\mathbf{r},t)$
is the mass density of MDW. As discussed below,
the mass density perturbations related to the recoil and MDW
effects become spatially well separated in the vicinity of the left
interface of the medium block in Fig.~\ref{fig:illustration}
after the light pulse has penetrated in the medium.

When the atoms are displaced from their equilibrium positions, they are also affected
by the elastic force density $\mathbf{f}_\mathrm{el}(\mathbf{r},t)$ .
As the atomic velocities are nonrelativistic,
we can apply Newtonian mechanics for the description
of the movement of atoms.
Newton's equation of motion for the mass density
of the medium is given by
\begin{equation}
 \rho_\mathrm{a}(\mathbf{r},t)\frac{d^2\boldsymbol{r}_\mathrm{a}(\mathbf{r},t)}{dt^2}=\mathbf{f}_\mathrm{opt}(\mathbf{r},t)+\mathbf{f}_\mathrm{el}(\mathbf{r},t),
 \label{eq:mediumnewton}
\end{equation}
where $\boldsymbol{r}_\mathrm{a}(\mathbf{r},t)$ is the position- and time-dependent atomic displacement
field of the medium. As we will see in numerical calculations, the first term on the right in Eq.~\eqref{eq:mediumnewton}
dominates the second term in the time scale of the light propagation (cf. Fig.~\ref{fig:mdw} below). However,
in longer time scales, the second term becomes important as
it relaxes the nonequilibrium of the mass density of the medium especially
near the material surfaces, which have received momentum
from the light field. In very precise calculations, it can be seen that
the elastic forces also affect at the time scale of light propagation,
but these effects are much smaller than the surface effects.

\subsubsection{Elastic forces}
In order to use Newton's equation of motion in Eq.~\eqref{eq:mediumnewton} to calculate
the dynamics of the medium, we need to have an expression for the elastic force density $\mathbf{f}_\mathrm{el}(\mathbf{r},t)$.
Close to equilibrium, the elastic forces between atoms are known to be well described by Hooke's law.
In the most simple case of a homogeneous isotropic elastic medium,
the stiffness tensor in Hooke's law has only two independent entries.
These entries are typically described by using the Lam\'e parameters
or any two independent elastic moduli, such as the bulk modulus $B$
and the shear modulus $G$ \cite{Mavko2003}.
In this case, the elastic force density
in terms of the material displacement field $\mathbf{r}_\mathrm{a}(\mathbf{r},t)$
is well known to be given by \cite{Bedford1994}
\begin{equation}
 \mathbf{f}_\mathrm{el}(\mathbf{r},t)=\textstyle(B+\frac{4}{3}G)\nabla[\nabla\cdot\mathbf{r}_\mathrm{a}(\mathbf{r},t)]-G\nabla\times[\nabla\times\mathbf{r}_\mathrm{a}(\mathbf{r},t)].
 \label{eq:elasticforcedensity}
\end{equation}
The factor of the first term $B+\frac{4}{3}G$ is also
known as the $P$-wave modulus \cite{Mavko2003}.
In the case of fluids, the shear modulus $G$ describes
dynamic viscosity. Therefore, in the special case
of non-viscous fluids, we could set $G=0$, when the number
of independent elastic moduli is reduced to one
and the second term of Eq.~\eqref{eq:elasticforcedensity}
becomes zero.

Note that the description of a fluid using the elastic force
in Eq.~\eqref{eq:elasticforcedensity} is only possible in the case
of small atomic displacements. The difference
between a solid and a fluid becomes apparent
in the case of larger atomic displacements
when one must also take convection
into account leading to Navier-Stokes equations.

\subsubsection{Displacement of atoms due to optical and elastic forces}

Newton's equation of motion in Eq.~\eqref{eq:mediumnewton} and the optical
and elastic force densities in Eqs.~\eqref{eq:opticalforcedensity} and \eqref{eq:elasticforcedensity}
can be used to calculate the position and velocity distributions of atoms in the medium
as a function of time. The total displacement of atoms at position $\mathbf{r}$ solved from Eq.~\eqref{eq:mediumnewton}
as a function of time is given by integration as
\begin{align}
 \mathbf{r}_\mathrm{a}(\mathbf{r},t) &=\int_{-\infty}^t\int_{-\infty}^{t''}\frac{d^2\boldsymbol{r}_\mathrm{a}(\mathbf{r},t')}{dt'^2}dt'dt''\nonumber\\
 &=\int_{-\infty}^t\int_{-\infty}^{t''}\frac{\mathbf{f}_\mathrm{opt}(\mathbf{r},t')+\mathbf{f}_\mathrm{el}(\mathbf{r},t')}{\rho_\mathrm{a}(\mathbf{r},t')}dt'dt''. 
 \label{eq:displacement}
\end{align}
As the atoms are very massive when compared to mass equivalent of the field energy,
the perturbed mass density of the medium $\rho_\mathrm{a}(\mathbf{r},t)$
is extremely close to the equilibrium mass density $\rho_0$. Therefore,
when applying Eq.~\eqref{eq:displacement}, it is well justified to
approximate the mass density in the denominator with the equilibrium
mass density $\rho_0$.

\subsubsection{Mass transferred by optoelastic forces}

Newton's equation of motion in Eq.~\eqref{eq:mediumnewton} can also be used
to calculate the mass transferred by MDW corresponding to Eq.~\eqref{eq:mdwmass1} in the MP quasiparticle model.
When the light pulse has passed through the medium, the displacement of atoms
is given by $\mathbf{r}_\mathrm{a}(\mathbf{r},\infty)$.
The displaced volume is given by
$\delta V=\int\mathbf{r}_\mathrm{a}(\mathbf{r},\infty)\cdot d\mathbf{A}$,
where the integration is performed over the transverse plane
with vector surface element $d\mathbf{A}$. By using the equation for the displacement
of atoms in Eq.~\eqref{eq:displacement}, for the total transferred mass
$\delta m=\rho_0\delta V$, we obtain an expression
\begin{equation}
 \delta m=\int\int_{-\infty}^\infty\int_{-\infty}^{t}[\mathbf{f}_\mathrm{opt}(\mathbf{r},t')+\mathbf{f}_\mathrm{el}(\mathbf{r},t')]dt'dt\cdot d\mathbf{A}.
 \label{eq:mdwmass2}
\end{equation}
Using the relation $\delta m=\int\rho_\text{\tiny MDW}(\mathbf{r},t)dV$
and $cdt=ndx$, we obtain the mass density of MDW, given by
\begin{equation}
 \rho_\text{\tiny MDW}(\mathbf{r},t)=\frac{n}{c}\int_{-\infty}^{t}[\mathbf{f}_\mathrm{opt}(\mathbf{r},t')+\mathbf{f}_\mathrm{el}(\mathbf{r},t')]\cdot\hat{\mathbf{x}}\,dt',
 \label{eq:mdwdensity}
\end{equation}
where $\hat{\mathbf{x}}$ is the unit vector in the direction of light propagation.
With proper expressions for the optical and elastic forces, Eq.~\eqref{eq:mdwdensity} can be used
for numerical simulations of the propagation of the light associated MDW in the medium.

\subsubsection{Momentum of the mass density wave}

Next we present an expression for the mechanical momentum of the MDW atoms
corresponding to Eq.~\eqref{eq:momentumsplitting} in the MP quasiparticle model.
Using the velocity distribution of the medium, given by
$\mathbf{v}_\mathrm{a}(\mathbf{r},t)=d\boldsymbol{r}_\mathrm{a}(\mathbf{r},t)/dt$,
the momentum of MDW is directly given by
integration of the classical momentum density $\rho_0\mathbf{v}_\mathrm{a}(\mathbf{r},t)$ as
\begin{equation}
 \mathbf{p}_\text{\tiny MDW}=\int \rho_0\mathbf{v}_\mathrm{a}(\mathbf{r},t)d^3r=\int \rho_\text{\tiny MDW}(\mathbf{r},t)\mathbf{v}d^3r,
 \label{eq:mdwmomentum}
\end{equation}
where $\mathbf{v}$ is the velocity vector of MP.
In numerical simulations in Sec.~\ref{sec:simulations},
we will verify that these expressions give an equal result,
which is also equal to the expression in terms of the
Poynting vector in Eq.~\eqref{eq:dipolemomentum}
and the expression in the MP quasiparticle model
in Eq.~\eqref{eq:momentumsplitting}.

\subsubsection{Kinetic energy of the mass density wave}

Another question of special interest is the accurate evaluation of the kinetic
energy of atoms in MDW, which was neglected in the MP quasiparticle model in Sec.~\ref{sec:mptheory}.
The density of kinetic energy is given by
$u_\mathrm{kin}(\mathbf{r},t)=\frac{1}{2}\rho_\mathrm{a}(\mathbf{r},t)\mathbf{v}_\mathrm{a}(\mathbf{r},t)^2$,
and the total kinetic energy is then given by an integral
$\Delta E_\mathrm{kin}=\int u_\mathrm{kin}(\mathbf{r},t)d^3r=\int \frac{1}{2}\rho_\mathrm{a}(\mathbf{r},t)\mathbf{v}_\mathrm{a}(\mathbf{r},t)^2 d^3r
=\int \frac{1}{2}\rho_\mathrm{a}(\mathbf{r},t)[d\boldsymbol{r}_\mathrm{a}(\mathbf{r},t)/dt]d^3r$.
Substituting to this expression the velocity field of the medium solved from
Eq.~\eqref{eq:mediumnewton} and approximating $\rho_\mathrm{a}(\mathbf{r},t)\approx\rho_0$
gives the kinetic energy of atoms in MDW as
\begin{align}
 \Delta E_\mathrm{kin} &=\int\frac{\rho_0}{2}\Big(\frac{\partial\mathbf{r}_\mathrm{a}(\mathbf{r},t)}{\partial t}\Big)^2d^3r\nonumber\\
 &=\int \frac{1}{2\rho_0}\Big[\int_{-\infty}^t[\mathbf{f}_\mathrm{opt}(\mathbf{r},t')+\mathbf{f}_\mathrm{el}(\mathbf{r},t')]dt'\Big]^2 d^3r.
 \label{eq:kineticenergy}
\end{align}
It can be seen that in the limit of an infinite mass density of the medium,
$\rho_0\longrightarrow\infty$, the kinetic energy of atoms in MDW becomes
zero, $\Delta E_\mathrm{kin}\longrightarrow 0$. This is the conventional approximation
in solid state electrodynamics, but it is not exactly correct.
The movement of atoms constituting MDW, in fact, presents a crucial part of the
covariant theory of light in a medium and gives rise to the increase of the momentum
of light from the vacuum value to the Minkowski value $p=n\hbar\omega/c$.

\subsubsection{Strain energy of the mass density wave}

The strain energy of MDW was neglected in the MP quasiparticle model in Sec.~\ref{sec:mptheory},
but here we take it into account.
After calculation of the atomic displacement field $\mathbf{r}_\mathrm{a}(\mathbf{r},t)$
by using Eq.~\eqref{eq:displacement}, it is a straightforward task to
evaluate the strain energy of MDW by integrating the standard form of the
elastic energy density given, e.g., in Ref.~\citenum{Kittel2005} as
\begin{align}
 \Delta E_\mathrm{strain} &=\int\!\!\Big\{\!\frac{B+\frac{4}{3}G}{2}\Big[\Big(\frac{\partial r_{\mathrm{a},x}}{\partial x}\Big)^2\!+\!\Big(\frac{\partial r_{\mathrm{a},y}}{\partial y}\Big)^2\!+\!\Big(\frac{\partial r_{\mathrm{a},z}}{\partial z}\Big)^2\Big]\nonumber\\
 &\hspace{0.5cm}+\frac{G}{2}\Big[\Big(\frac{\partial r_{\mathrm{a},x}}{\partial y}+\frac{\partial r_{\mathrm{a},y}}{\partial x}\Big)^2\!+\!\Big(\frac{\partial r_{\mathrm{a},x}}{\partial z}+\frac{\partial r_{\mathrm{a},z}}{\partial x}\Big)^2\nonumber\\
 &\hspace{0.5cm}+\Big(\frac{\partial r_{\mathrm{a},y}}{\partial z}+\frac{\partial r_{\mathrm{a},z}}{\partial y}\Big)^2\Big]
 +(B-{\textstyle\frac{2}{3}}G)\Big[\frac{r_{\mathrm{a},x}}{dx}\frac{r_{\mathrm{a},y}}{dy}\nonumber\\
 &\hspace{0.5cm}+\frac{r_{\mathrm{a},x}}{dx}\frac{r_{\mathrm{a},z}}{dz}+\frac{r_{\mathrm{a},y}}{dy}\frac{r_{\mathrm{a},z}}{dz}\Big]\Big\}d^3r,
 \label{eq:strainenergy}
\end{align}
where $r_{\mathrm{a},x}$, $r_{\mathrm{a},y}$, and $r_{\mathrm{a},z}$
correspond to the $x$-, $y$-, and $z$-components of the material displacement field
$\mathbf{r}_\mathrm{a}(\mathbf{r},t)$.

\subsubsection{Recoil energy at the surface of the medium}

The recoil energy at the surface of the medium, resulting from the optical surface force density in the first
term of Eq.~\eqref{eq:opticalforcedensity}, consists of both the kinetic energy
of surface atoms and the elastic potential energy between the atoms at the surface.
It depends on the thickness of the surface layer that takes the recoil momentum.
The kinetic energy contribution can be calculated by using the same kinetic energy
formula as the kinetic energy of MDW in Eq.~\eqref{eq:kineticenergy} and
the elastic potential energy of surface atoms can be calculated
by using the strain energy in Eq.~\eqref{eq:strainenergy}.

\subsubsection{Elastic wave equation}
After the light pulse has escaped the medium, the atoms have been left
displaced from their equilibrium positions and they continue to interact through elastic forces. Therefore,
following from Eqs.~\eqref{eq:mediumnewton} and \eqref{eq:elasticforcedensity},
the atomic displacement field of the medium
$\mathbf{r}_\mathrm{a}(\mathbf{r},t)$, in the absence of optical forces,
obeys the elastic wave equation, given by \cite{Bedford1994}
\begin{equation}
 \frac{d^2\boldsymbol{r}_\mathrm{a}(\mathbf{r},t)}{dt^2}=v_\parallel^2\nabla[\nabla\cdot\mathbf{r}_\mathrm{a}(\mathbf{r},t)]-v_\perp^2\nabla\times[\nabla\times\mathbf{r}_\mathrm{a}(\mathbf{r},t)].
 \label{eq:elasticwaveequation}
\end{equation}
Here $v_\parallel=\sqrt{(B+\frac{4}{3}G)/\rho_0}$ is the velocity of longitudinal compressional waves and
$v_\perp=\sqrt{G/\rho_0}$ is the velocity of transverse shear waves.
Therefore, we expect that
after the formation of the mass density perturbations as a result of the
optoelastic forces, we are likely to observe the elastic relaxation
of the mass nonequilibrium of the medium at the sound velocities $v_\parallel$ and $v_\perp$.
In the numerical simulations presented in Sec.~\ref{sec:simulations},
we will see that the elastic relaxation process in fact takes place.

\section{\label{sec:simulations}Simulations of the mass transfer}

Above, we have derived the rest mass, momentum, and the transferred mass of
MP using the conservation laws and the covariance principle in Sec.~\ref{sec:mptheory}.
Independently, expressions for the momentum and the transferred mass of MP
were derived using the OCD approach based on the semiclassical electrodynamics and the
continuum mechanics in Sec.~\ref{sec:continuum}.
Next, we present numerical simulations on the MDW and recoil effects as well as the relaxation
dynamics of the mass nonequilibrium resulting from the mass transfer.

\begin{figure}
\centering
\includegraphics[width=0.48\textwidth]{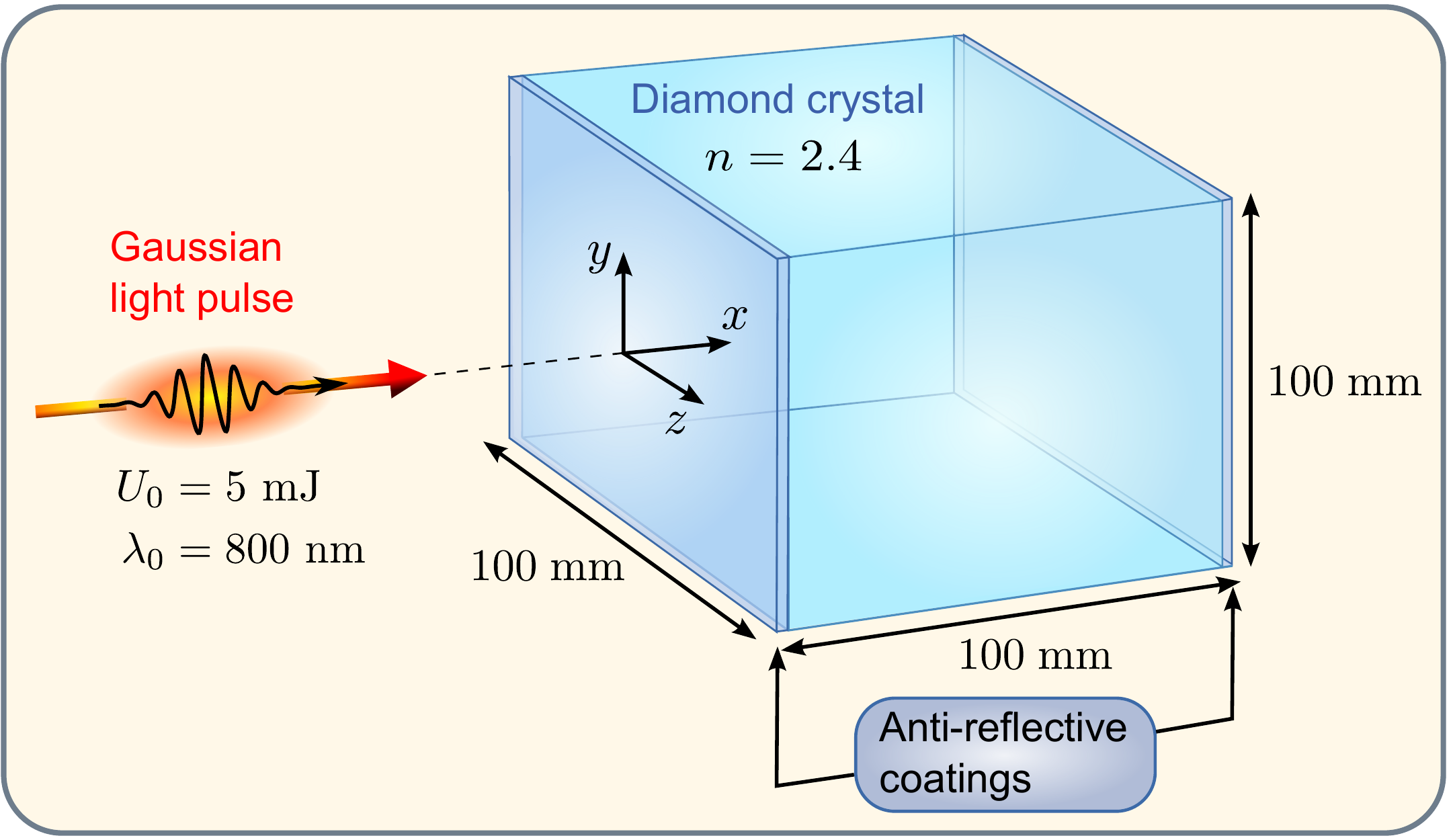}
\caption{\label{fig:simulationgeometry}
(Color online) The geometry of the simulations consisting of a cubic diamond crystal block coated
with anti-reflective coatings. The refractive index of the crystal is $n=2.4$.
A Gaussian light pulse of energy $U_0=5$ mJ and central wavelength $\lambda_0=800$ nm
propagates in the direction of the positive $x$ axis and enters the crystal from the left.
The geometry is centered so that the center of the light pulse enters the crystal at $x=y=z=0$ mm.
The first interface of the crystal is located at $x=0$ mm and the second
interface at $x=100$ mm.}
\vspace{-0.2cm}
\end{figure}

The simulations are done for a diamond crystal with the refractive
index $n=2.4$ \cite{Phillip1964}, mass density $\rho_0=3500$ kg/m$^3$ \cite{Lide2004},
bulk modulus $B=443$ GPa \cite{Kittel2005}, and shear modulus $G=478$ GPa \cite{McSkimin1972}.
The simulation geometry of a cubic diamond crystal block
with anti-reflective coatings is illustrated in Fig.~\ref{fig:simulationgeometry}.
The first and second interfaces of the crystal are located at positions
$x=0$ and $100$ mm. In the $y$ and $z$ directions, the geometry is centered so
that the trajectory of the light pulse follows the line $y=z=0$ mm.

We assume a titanium-sapphire laser
pulse with a wavelength $\lambda_0=800$ nm ($\hbar\omega_0=1.55$ eV) and
total energy $U_0=5$ mJ. This corresponds to the photon number
of $N_0=U_0/\hbar\omega_0=2.0\times 10^{16}$. The Gaussian form of the electromagnetic
wave packet is assumed to propagate in the $x$ direction.
For the mathematical description of the electric and magnetic fields
of the pulse in the one- and three-dimensional simulations, see
Secs.~\ref{sec:1dsimulations} and \ref{sec:3dsimulations}, where
we also describe the corresponding simulations.
For the flowchart and other computational details of the
simulations, see Appendix \ref{apx:flowchart}.

\subsection{\label{sec:1dsimulations}Simulations in one dimension}

In the one-dimensional simulations, we consider a plate that has thickness
$L=100$ mm in the $x$ direction and is infinite in the $y$ and $z$ directions.
The light pulse propagates along the $x$ axis which is also one of
the principal axes of the single crystal.
As exact solutions to the Maxwell's equations,
the electric and magnetic fields of the one-dimensional Gaussian pulse, with energy
$U_0$ per cross-sectional area $A$, are given by \cite{Griffiths1998}
\begin{align}
 \mathbf{E}(\mathbf{r},t) & =\mathrm{Re}\Big[\int_{-\infty}^\infty \tilde E(k)e^{i(kx-\omega(k) t)}dk\Big]\hat{\mathbf{y}}\nonumber\\ 
 & =\sqrt{\frac{2n\Delta k_x U_0}{\pi^{1/2}\varepsilon A(1+e^{-(k_0/\Delta k_x)^2})}}\nonumber\\
 &\hspace{0.5cm}\times\cos\!\Big(nk_0(x-ct/n)\Big) e^{-(n\Delta k_x)^2(x-ct/n)^2/2}\hat{\mathbf{y}},
 \label{eq:electricfield1d}
\end{align}
\vspace{-0.2cm}
\begin{align}
 \mathbf{H}(\mathbf{r},t) & =\mathrm{Re}\Big[\int_{-\infty}^\infty \tilde H(k)e^{i(kx-\omega(k) t)}dk\Big]\hat{\mathbf{z}}\nonumber\\ 
 &=\sqrt{\frac{2n\Delta k_x U_0}{\pi^{1/2}\mu A(1+e^{-(k_0/\Delta k_x)^2})}}\nonumber\\
 &\hspace{0.5cm}\times\cos\!\Big(nk_0(x-ct/n)\Big) e^{-(n\Delta k_x)^2(x-ct/n)^2/2}\hat{\mathbf{z}}.
 \label{eq:magneticfield1d}
\end{align}
Here $\tilde E(k)=\tilde E_0e^{-[(k-nk_0)/(n\Delta k_x)]^2/2}$
and $\tilde H(k)=\tilde H_0e^{-[(k-nk_0)/(n\Delta k_x)]^2/2}$ are Gaussian functions
in which $\tilde E_0$ and $\tilde H_0$ are normalization factors,
$\omega(k)=ck/n$ is the dispersion relation of a nondispersive medium,
$k_0=\omega_0/c$ is the wave number corresponding to the central
frequency $\omega_0$ in vacuum and $\Delta k_x$ is the standard deviation
of the wave number in vacuum. It defines the pulse width
in the $x$ direction. We assume that the relative spectral width
of the pulse, in our example, is $\Delta\omega/\omega_0=\Delta k_x/k_0=10^{-5}$.
Then, we have $\Delta k_x=10^{-5}k_0$ and the corresponding standard
deviation of position in vacuum is
$\Delta x=1/(\sqrt{2}\Delta k_x)\approx 9$ mm,
which is much shorter than the crystal block. The standard deviation of the
pulse width in time is then $\Delta t=\Delta x/c\approx 30$ ps.
The normalization factors in Eqs.~\eqref{eq:electricfield1d}
and \eqref{eq:magneticfield1d} have been determined
so that the integral of the corresponding instantaneous
energy density over $x$ gives $U_0/A$.

In the simulations, for the pulse energy $U_0=5$ mJ, we use the cross-sectional
area given by $A=(\lambda/2)^2$,
where $\lambda=\lambda_0/n$ is the wavelength in the crystal.
The seemingly high power per unit area was chosen so that we can obtain
an \emph{order-of-magnitude} estimate of how large atomic displacements
we obtain if the whole vacuum energy $U_0=5$ mJ of the laser pulse can be coupled to a
free-standing waveguide having a cross section $(\lambda/2)^2$ (see discussion
in Sec.~\ref{sec:measurements}).

\begin{figure}
\centering
\includegraphics[width=\columnwidth]{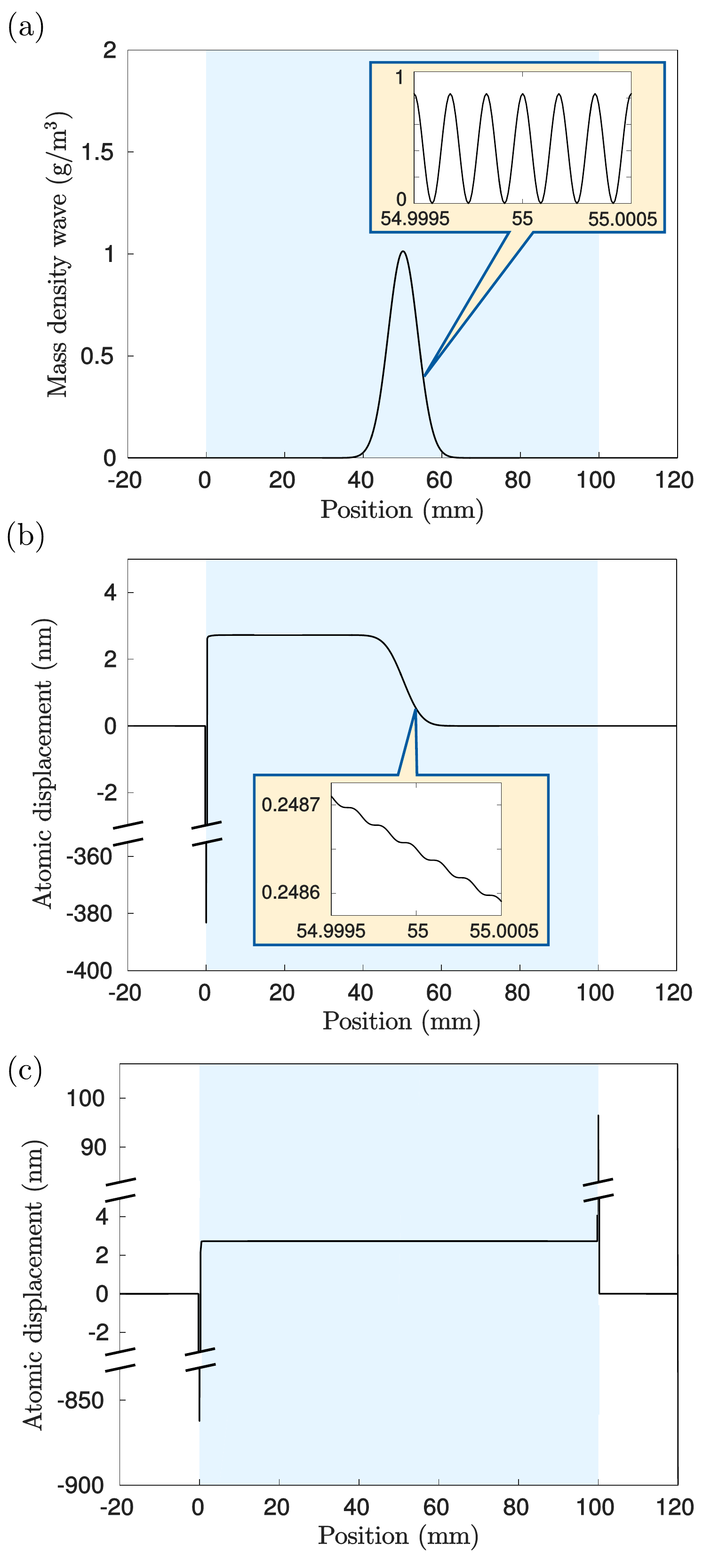}
\caption{\label{fig:mdw}
(Color online) Simulation of the mass transfer
due to a Gaussian light pulse in one dimension.
(a) The calculated mass density of MDW, i.e., the excess mass density in the medium, as a function
of position when the light pulse is in the middle of the crystal.
The light blue background represents the region of the crystal between $x=0$ and $100$ mm.
The focused subgraph shows the exact instantaneous MDW near $x=55$ mm.
(b) The calculated atomic displacements when the light pulse is in the middle of the crystal.
The focused subgraph shows the exact instantaneous atomic displacements near $x=55$ mm.
(c) The calculated atomic displacements when the light pulse has just left the crystal.
Note the breaks in the scales of the figures.}
\end{figure}

We perform the one-dimensional mass transfer simulations in two methods:
(1) First, we use the exact instantaneous electric and magnetic fields given
in Eqs.~\eqref{eq:electricfield1d} and \eqref{eq:magneticfield1d}
for the calculation of the optical force density in Eq.~\eqref{eq:opticalforcedensity}.
In this simulation, we use a fine discretization with $h_x=\lambda/40$ and $h_t=2\pi/(40\omega_0)$
that is sufficiently dense compared to the scale of the harmonic cycle.
(2) Second, in order to justify the approximations made to speed up
the three-dimensional simulations, we also perform the one-dimensional
simulations by using the following approximation.
We average the two terms of the optical force density in Eq.~\eqref{eq:opticalforcedensity}
over the harmonic cycle assuming that the exponential time-dependent factor
of the fields in Eqs.~\eqref{eq:electricfield1d} and \eqref{eq:magneticfield1d}
is constant over the harmonic cycle. This approximation allows
us to use a very coarse \emph{time and space}
discretization in our simulations
when compared to the scale of the harmonic cycle.
The spatial and temporal discretization lengths in
this simulation are $h_x=250$ $\mu$m and $h_t=1$ ps,
which are small compared to the spatial and temporal widths of the pulse.

Within numerical accuracy of the coarse-grid computations,
the total transferred mass and momentum computed by using the
time-averaged force density and the coarse grid (method 2)
are found to be equal to those obtained by using
the exact instantaneous force density (method 1).
This result justifies the
use of the coarse grid in the three-dimensional
simulations described in Sec.~\ref{sec:3dsimulations}.

As described in Appendix \ref{apx:flowchart},
in the present simulations, we have described the refractive index
as a step function near the surfaces, thus neglecting any atomic scale changes in it.
However, in the quantitative calculation of the atomic displacements near the material
interfaces resulting from the optical surface force density described by the first term
in Eq.~\eqref{eq:opticalforcedensity}, this approximation should be considered very
carefully. The same applies for the relaxation of the mass nonequilibrium by elastic waves
that has been described in the case of the three-dimensional simulations in Sec.~\ref{sec:3dsimulations}.
Therefore, our calculations related to these quantities should be considered only
approximative.

Figure \ref{fig:mdw}(a) shows the simulation of
MDW as a function of position
when the light pulse is propagating in the middle of the crystal.
This graph is obtained from the simulation
with a coarse grid and time-averaging over the harmonic cycle.
MDW equals the difference of the disturbed mass density $\rho_\mathrm{a}(\mathbf{r},t)$
and the equilibrium mass density $\rho_0$
inside the crystal and it is calculated by using Eq.~\eqref{eq:mdwdensity}.
MDW is driven by the optoelastic forces due to the Gaussian
light pulse. The mass density perturbance at the first interface due to the
interface force is not shown in this panel.
The form of MDW clearly follows the Gaussian
form of the pulse as expected. When we integrate the MDW mass density
in Fig.~\ref{fig:mdw}(a), we obtain the total
transferred mass of $2.6\times 10^{-19}$ kg. Dividing this by
the photon number of the light pulse, we then obtain the value of
the transferred mass per photon, given by $7.4$ eV/$c^2$.
This corresponds to the value obtained in the MP
quasiparticle approach by using Eq.~\eqref{eq:mdwmass1}.
The subgraph focused near $x=55$ mm
in Fig.~\ref{fig:mdw}(a)
shows the actual functional form of MDW obtained by using
instantaneous fields and the fine discretization.
The Gaussian envelope of the pulse cannot be seen in this scale.

Figure \ref{fig:mdw}(b) shows the atomic displacements corresponding
to MDW in Fig.~\ref{fig:mdw}(a), again,
as a function of position.
On the left from $x=0$ mm, the atomic
displacement is zero as there are no atoms in vacuum, where the
refractive index is unity.
Due to the optoelastic recoil effect described by
the first term of Eq.~\eqref{eq:opticalforcedensity},
a thin material layer at the interface recoils to the left.
Therefore, the atomic displacement at the interface is negative.
The atomic displacement has a constant value of 
of 2.7 nm between positions 0 and 40 mm. This follows from
the optical force in the second term of Eq.~\eqref{eq:opticalforcedensity}.
In Fig.~\ref{fig:mdw}(b), the leading edge of the
optical pulse is propagating to the right approximately at the position
$x=60$ mm. Therefore, to the right of $x=60$ mm, the atomic displacement is zero.
The optoelastically driven MDW is manifested by the fact that atoms
are more densely spaced at the position of the light pulse as the atoms
on the left of the pulse have been displaced forward and the atoms
on the right of the pulse are still at their equilibrium positions.
The momentum of atoms in MDW is obtained by 
integrating the classical momentum density as given in Eq.~\eqref{eq:mdwmomentum}
at an arbitrary time when the entire light pulse is contained in the medium.
The calculation verifies Eq.~\eqref{eq:momentumsplitting} of the MP
quasiparticle model within the precision of the numerical accuracy of the simulation.
Therefore, the numerical OCD simulations are seen to be fully
consistent with the MP quasiparticle approach
described in Sec.~\ref{sec:mptheory}.
The subgraph focused near the position
$x=55$ mm in Fig.~\ref{fig:mdw}(b)
shows the atomic displacements computed using
instantaneous fields and a fine discretization.
The effect of the variation of the optical force density within the harmonic
cycle is clearly visible in this instantaneous atomic displacement.

Figure \ref{fig:mdw}(c) shows the atomic displacements just after the light pulse
has left the medium. One can see that all atoms inside the crystal have been displaced
forward from their initial positions. The surface atoms at the both surfaces
have been displaced outwards from the medium due to the optoelastic recoil effect.
The magnitudes of the atomic displacements at the interfaces are changing
as a function of time due to the elastic forces, which, after the pulse 
transmission, start to restore the mass density equilibrium in the crystal.
When the equilibrium has been reestablished, the elastic energy that was
left in the crystal after the light pulse is converted to lattice heat.
The relaxation of the mass nonequilibrium is studied in more detail
in the case of the three-dimensional simulations in Sec.~\ref{sec:3dsimulations}.

\subsection{\label{sec:3dsimulations}Simulations in three dimensions}

Next we present corresponding simulations in a full three-dimensional geometry in Fig.~\ref{fig:simulationgeometry}
and study also the relaxation of the mass nonequilibrium resulting from the
MDW and recoil effects.
We have included the three-dimensional simulations in our work since
the three-dimensional model gives a deeper insight to the strain fields and their relaxation by
sound waves. In the simulation, the medium is discretized to cubic voxels with an edge
width $h_x=h_y=h_z=250$ $\mu$m.
As approximate solutions to the Maxwell's equations,
the electric and magnetic fields of the three-dimensional Gaussian pulse are given by
\begin{align}
 &\mathbf{E}(\mathbf{r},t)\nonumber\\
 &=\sqrt{\frac{2n\Delta k_x\Delta k_y\Delta k_z U_0}{\pi^{3/2}\varepsilon(1+e^{-(k_0/\Delta k_x)^2})}}
 \,\cos\!\Big(nk_0(x-ct/n)\Big)\nonumber\\
 &\hspace{0.5cm}\times e^{-(n\Delta k_x)^2(x-ct/n)^2/2}e^{-(\Delta k_y)^2y^2/2}e^{-(\Delta k_z)^2z^2/2}\hat{\mathbf{y}},
 \label{eq:electricfield3d}
\end{align}
\vspace{-0.2cm}
\begin{align}
 &\mathbf{H}(\mathbf{r},t)\nonumber\\
 &=\sqrt{\frac{2n\Delta k_x\Delta k_y\Delta k_z U_0}{\pi^{3/2}\mu(1+e^{-(k_0/\Delta k_x)^2})}}
 \,\cos\!\Big(nk_0(x-ct/n)\Big)\nonumber\\
 &\hspace{0.5cm}\times e^{-(n\Delta k_x)^2(x-ct/n)^2/2}e^{-(\Delta k_y)^2y^2/2}e^{-(\Delta k_z)^2z^2/2}\hat{\mathbf{z}}.
 \label{eq:magneticfield3d}
\end{align}
Here $\Delta k_y$ and $\Delta k_z$ are the standard deviations
of the $y$ and $z$ components of the wave vector in vacuum. They define the pulse width
in the transverse plane. We use $\Delta k_y=\Delta k_z=10^{-4} k_0$.
This corresponds to the standard deviation of position
of $\Delta y=\Delta z\approx 0.9$ mm.
For other parameters, we use the same values
as in the case of the one-dimensional simulations in Sec.~\ref{sec:1dsimulations}.
The normalization factors in Eqs.~\eqref{eq:electricfield3d} and \eqref{eq:magneticfield3d}
are determined so that the volume integral
of the corresponding instantaneous energy density over
the light pulse gives $U_0$.

The approximate solutions of the electric and magnetic fields in Eqs.~\eqref{eq:electricfield3d}
and \eqref{eq:magneticfield3d} form an exact solution for $y=z=0$ mm.
They are also exact in the plane wave limit $\Delta k_y\longrightarrow 0$ and $\Delta k_z\longrightarrow 0$.
Therefore, as in our case $\Delta k_y$ and $\Delta k_z$ are sufficiently small,
we can consider our approximation as accurate. Quantitatively, the accuracy of the approximation is justified
by calculating the integral of the energy density corresponding to the longitudinal component of the magnetic
field that results from the exact application of Faraday's law to the electric field in Eq.~\eqref{eq:electricfield3d}
in vacuum. The ratio of this energy to the energy of the transverse component of the
magnetic field in Eq.~\eqref{eq:magneticfield3d} is found to be $5\times 10^{-9}$, which is
very small and hence justifies our approximation.
Note that the three- and one-dimensional simulations give the same value
for the momentum and the transferred mass of MDW per unit energy.
Since the one-dimensional Gaussian is an exact solution of the Maxwell's equations,
this proves the overall consistency of the approximative three-dimensional
Gaussian pulse used in the simulations.

To save computing power in the demanding
three-dimensional simulations, we average the optical force density in
Eq.~\eqref{eq:opticalforcedensity} over the harmonic cycle.
Since the pulse length in the time domain is much larger than the harmonic cycle,
the exponential time dependent factor in the fields in Eqs.~\eqref{eq:electricfield3d}
\eqref{eq:magneticfield3d} can be considered to be approximately constant over the harmonic cycle.
The Poynting vector $\mathbf{S}=\mathbf{E}\times\mathbf{H}$
averaged over the harmonic cycle is then given by
\begin{align}
 \langle\mathbf{S}(\mathbf{r},t)\rangle &=\frac{U_0c\Delta k_x\Delta k_y\Delta k_z}{\pi^{3/2}(1+e^{-(k_0/\Delta k_x)^2})}\,e^{-(n\Delta k_x)^2(x-ct/n)^2}\nonumber\\
 &\hspace{0.5cm}\times e^{-(\Delta k_y)^2y^2}e^{-(\Delta k_z)^2z^2}\hat{\mathbf{x}}.
 \label{eq:Gaussianpoynting}
\end{align}
The optical force without interface terms is proportional to the time derivative
of the averaged Poynting vector in Eq.~\eqref{eq:Gaussianpoynting}.
This approximation is justified by the one-dimensional calculations
described in Sec.~\ref{sec:1dsimulations},
where we compared the accuracy of the solution obtained by using
a fine grid with instantaneous force density to
the accuracy of the solution obtained by using a coarse grid with
force density averaged over the harmonic cycle.
Also note that, due to not accounting for any near interface dependence
of the refractive index, our calculations regarding the atomic
displacements near the material interfaces should be considered only
approximative as described in more detail in Appendix \ref{apx:flowchart}.

Figure \ref{fig:mdwsimulation} shows the simulation of
MDW as a function of position at time $t=340$ ps in the plane $z=0$ mm.
The whole simulation is presented as a video file in the Supplemental Material \cite{supplementaryvideo}.
The simulation results are again fully consistent with the
MP quasiparticle model presented in Sec.~\ref{sec:mptheory}
as the total transferred mass, i.e., the integral of the mass density
in Fig.~\ref{fig:mdwsimulation}, very accurately equals the result
in Eq.~\eqref{eq:mdwmass1}, again, within
the numerical accuracy of the simulation.

\begin{figure}
\centering
\includegraphics[width=0.48\textwidth]{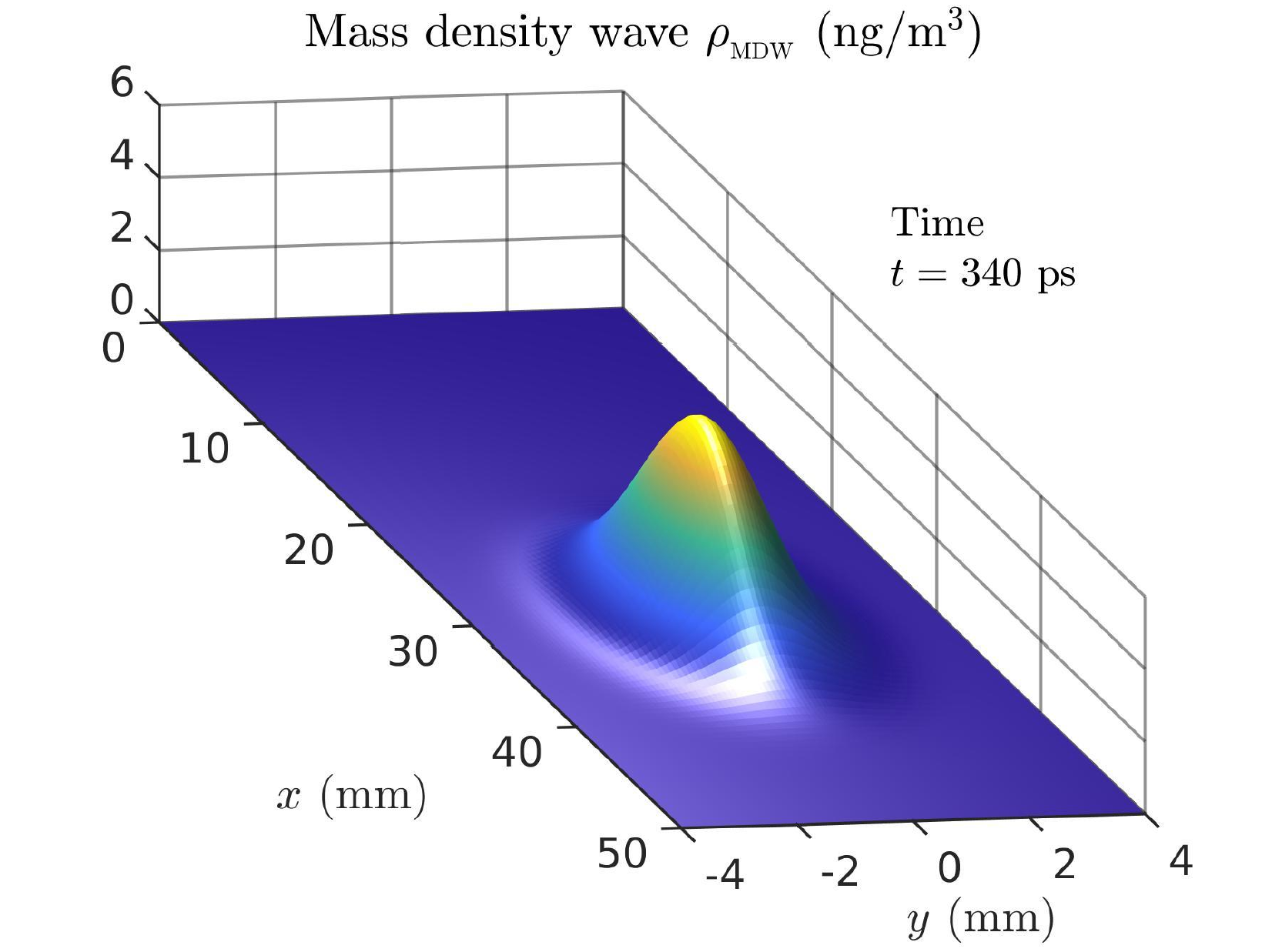}
\caption{\label{fig:mdwsimulation}
(Color online) Simulation of MDW driven by optoelastic forces.
The difference of the actual mass density of the medium $\rho_\mathrm{a}(\mathbf{r},t)$ and the equilibrium
mass density $\rho_0$ was plotted as a function of time and position in the plane $z=0$ mm.
The figure presents the position distribution at time $t=340$ ps after the start of the simulation.
The light pulse propagates in the direction of the positive $x$ axis. 
The front end of the crystal is located at $x=0$ mm and the back end at $x=100$ mm, not
shown in this figure. The mass density disturbance at the front interface is not drawn.}
\vspace{-0.2cm}
\end{figure}

\begin{figure*}
\centering
\includegraphics[width=\textwidth]{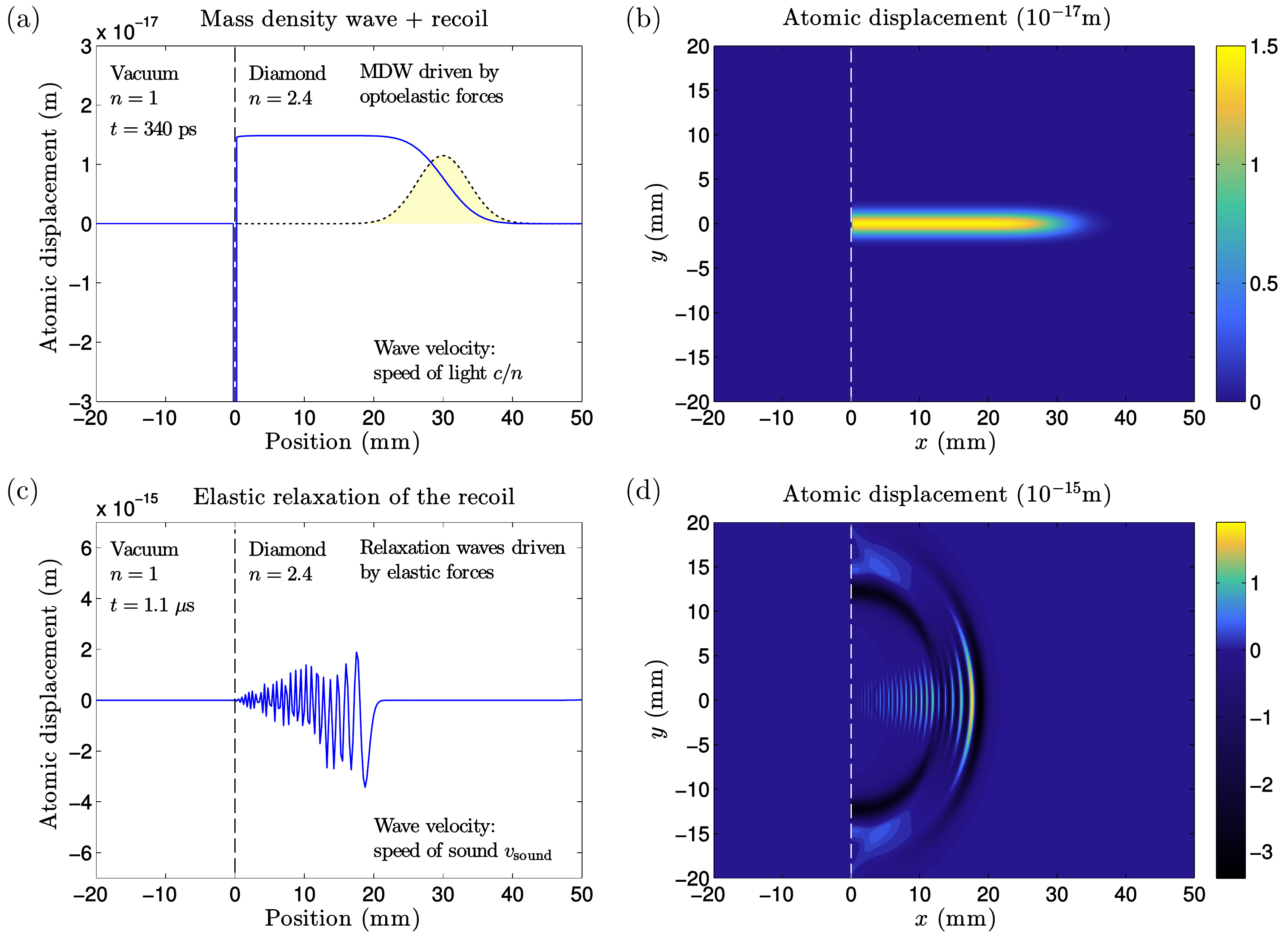}
\caption{\label{fig:displacementsimulation}
(Color online) Simulation of the atomic displacements due to optical
and elastic forces as a Gaussian light pulse enters from vacuum to a diamond crystal.
(a) The $x$ component of the atomic displacements plotted as a function
of $x$ for $y=z=0$ mm at time $t=340$ ps. The optoelastic forces drive MDW to the right
with velocity $v=c/n$ near $x=30$ mm. The dotted line represents the position of the Gaussian
light pulse driving MDW. Atoms in a thin interface layer at $x=0$ mm recoil to the left.
(b) The corresponding atomic displacements plotted in the plane $z=0$ mm.
(c) The same simulation at a later time $t=1.1$ $\mu$s. 
The light pulse has gone and the recoil of the interface atoms starts to relax.
Elastic forces between atoms drive density waves that propagate
to the right at the velocity of sound.
Due to the very approximative treatment of the near-interface region, only
the order of magnitude of atomic displacements and the position of the first wavefront
are physically significant.
(d) The corresponding atomic displacements plotted in the plane $z=0$ mm.
The dashed line represents the position of
the front vacuum-diamond interface. The back end of the crystal
is located at $x=100$ mm, not shown in this figure.
Therefore, in (c) and (d), we see only the relaxation transient
close to the front end of the crystal.}
\end{figure*}

Figure \ref{fig:displacementsimulation} shows the simulation
of the atomic displacements as a function of
position along $x$ axis for fixed $y=z=0$ mm. 
The whole simulation is presented as a video file in Supplemental Material
\cite{supplementaryvideo}.
Figure \ref{fig:displacementsimulation}(a)
presents the $x$-component of the atomic displacements at time $t=340$ ps.
This snapshot is taken in the time scale relevant
for describing the propagation of the Gaussian light pulse and,
at this moment, the maximum of the Gaussian pulse is at the position
$x=30$ mm. Figure \ref{fig:displacementsimulation}(a)
clearly corresponds to Fig.~\ref{fig:mdw}(b)
of the one-dimensional simulation.
Now, the magnitude of the atomic displacement at the interface
is of the order of $10^{-15}$ m,
which is not shown in the scale of the figure,
and the constant atomic displacement
after the light pulse is $1.5\times 10^{-17}$ m.

Figure \ref{fig:displacementsimulation}(b) shows
the atomic displacements
corresponding to Fig.~\ref{fig:displacementsimulation}(a)
as a function of $x$ and $y$ in the plane $z=0$ mm.
In the $y$ direction, for increasing and decreasing
values of $y$, the atomic displacement reaches zero
as the light-matter interaction takes place only
in the region of the light pulse and the elastic forces
are not fast enough to displace atoms in this short
time scale. The negative
atomic displacement at the interface
is not shown in the scale of
Fig.~\ref{fig:displacementsimulation}(b).

A second snapshot of the atomic displacements is given in Fig.~\ref{fig:displacementsimulation}(c)
at time $t=1.1$ $\mu$s. At this moment, the light pulse has left the crystal
and the atoms have obtained, excluding the crystal interfaces,
a constant displacement along the $x$ axis for fixed $y=z=0$ mm.
After the light pulse has left the crystal, the atomic density
in the vicinity of the interfaces is still different from the equilibrium value
and starts to relax through elastic waves as demonstrated
in Fig.~\ref{fig:displacementsimulation}(c).
Note that the scale of Fig.~\ref{fig:displacementsimulation}(c)
is chosen so that we can see the atomic displacements, which resulted from
the interface forces.
The positive constant atomic displacement seen in
Fig.~\ref{fig:displacementsimulation}(a) is very small and not visible in this scale.
In Fig.~\ref{fig:displacementsimulation}(c), we can see that the atomic displacements
at the interface are being relaxed by density waves driven by elastic forces.
Since the second interface of the crystal is at $x=100$ mm, we see only
the relaxation transient close to the first interface of the crystal.
The elastic wave is propagating to the right at the velocity of sound as expected.
The relaxation by elastic waves is governed by
the elastic wave equation in Eq.~\eqref{eq:elasticwaveequation}.
As the atomic displacements near the material interfaces
are computed only approximatively in our simulations due to the coarse grid,
the functional form of the ripples in the elastic waves in
Fig.~\ref{fig:displacementsimulation}(c) is not meaningful. Instead, only the
order of magnitude of the atomic displacements and the position of the
first wavefront as a function of time are physically meaningful in
Fig.~\ref{fig:displacementsimulation}(c).

Figure \ref{fig:displacementsimulation}(d) shows
the atomic displacements corresponding to Fig.~\ref{fig:displacementsimulation}(c)
as a function of $x$ and $y$ in the plane $z=0$ mm.
The same plot as a time-dependent
simulation can be found in the Supplemental Material \cite{supplementaryvideo}.
One can clearly see the wavefronts of the elastic waves.
The first of the two wide wavefronts with negative atomic displacements propagates at the
compressional wave velocity $v_\parallel$
while the second of the two wide wavefronts propagates
with the shear wave velocity $v_\perp$.
As in the case of Fig.~\ref{fig:displacementsimulation}(c),
the functional form of the ripples that exist in the elastic waves
especially near the line $y=z=0$ mm is not physically meaningful.

When the light pulse has propagated through the medium,
the MDW and recoil effects have displaced
medium atoms from their original positions.
Since the relaxation of the resulting mass nonequilibrium
takes place mainly by elastic processes as shown in the simulation
in Figs.~\ref{fig:displacementsimulation}(c) and
\ref{fig:displacementsimulation}(d), the equilibrium is achieved mainly
after the light pulse has escaped from the medium.
In the simulation, using Eqs.~\eqref{eq:kineticenergy} and
\eqref{eq:strainenergy}, we obtain the energy loss of $\Delta E_\mathrm{rec}=1.9\times 10^{-14}$ eV
due to the recoil effect at the two surfaces.
The recoil effect
is inversely proportional to the thickness $\Delta L$ of the medium layer at the interfaces which take the recoil.
The above energy loss was calculated for the relatively large value $\Delta L=250$ $\mu$m.
For the smaller value of $\Delta L$ given by the wavelength in the medium as
$\Delta L=\lambda=333$ nm, we obtain the recoil energy of $\Delta E_\mathrm{rec}=1.4\times 10^{-11}$ eV.
In any case, the recoil losses are seen to be very small.
The kinetic energy of MDW, calculated by using Eq.~\eqref{eq:kineticenergy},
is $\Delta E_\mathrm{kin}=3.6\times 10^{-16}$ eV and the strain energy of MDW, given by Eq.~\eqref{eq:strainenergy}, 
is $\Delta E_\mathrm{strain}=9.2\times 10^{-22}$ eV. Only a small fraction of these
energies is lost to heat so the recoil effect
is the main source of dissipation.
Note that, although the elastic energy of MDW
is small in our example, the effect of MDW on the momentum
of the light pulse is dramatic. The momentum of atoms
in MDW set in motion by the light pulse carry,
in our example, 83\% of the total momentum of the light pulse
[cf.~Eq.~\eqref{eq:momentumsplitting}].

Our theory can also be used to estimate
the effective imaginary part of the refractive index $n_\mathrm{i}$
resulting from the dissipation of the field energy
when the field propagates inside the medium.
We first compute the kinetic and strain energies left to the medium atoms per unit distance
traveled by light. We integrate the kinetic and strain energy densities, i.e., integrands
of Eqs.~\eqref{eq:kineticenergy} and \eqref{eq:strainenergy}, over the transverse
plane just after the light pulse has passed this plane.
Dividing this result with the field energy $U_0$ of the light pulse
gives the attenuation coefficient $1.6\times 10^{-36}$ 1/m.
Comparing this with the conventional expression $\alpha=2n_\mathrm{i}k_0$
for the attenuation coefficient, we obtain $n_\mathrm{i}=1.0\times 10^{-43}$.
This is vastly smaller than the imaginary part of the refractive index due
to other physical nonidealities in a highly transparent real material.
Therefore, the effect of $n_\mathrm{i}$
coming from the dissipation related to MDW on the optical
force density in Eq.~\eqref{eq:opticalforcedensity}
that is used to calculate the dynamics of the medium
is exceedingly small.

\section{\label{sec:comparison}Comparison of the results from the MP and OCD methods}

For the light pulse energy of $U_0=5$ mJ,
we obtain the photon number $N_0=2.0\times 10^{16}$.
When the transferred mass and the momentum of MP are multiplied
by $N_0$, we can directly compare the results of the MP and OCD models.
Using Eqs.~\eqref{eq:mdwmass1} and \eqref{eq:results}, we then obtain
\begin{equation}
 \delta m=\int \rho_\text{\tiny MDW}(\mathbf{r},t)d^3r=(n^2-1)N_0\hbar\omega/c^2,
\end{equation}
\begin{equation}
 \mathbf{p}_\text{\tiny MP}=\int \rho_0\mathbf{v}_\mathrm{a}(\mathbf{r},t)d^3r+\int \frac{\mathbf{S}(\mathbf{r},t)}{c^2}d^3r=\frac{nN_0\hbar\omega}{c}\hat{\mathbf{x}}.
\end{equation}
Comparison with the values obtained from the OCD simulations (left) and
the MP model (right) shows that they agree within the numerical accuracy of
the simulations. One may argue that this is as expected since the Lorentz invariance
is built in the Maxwell's equations and, therefore, they automatically account for the
covariance in the OCD approach. When we share the momentum in the OCD and MP models
into parts carried by the field and MDW, we correspondingly obtain the equalities 	
\begin{equation}
 \mathbf{p}_\text{\tiny MDW}=\int \rho_0\mathbf{v}_\mathrm{a}(\mathbf{r},t)d^3r=\Big(n-\frac{1}{n}\Big)\frac{N_0\hbar\omega}{c}\hat{\mathbf{x}},
\end{equation}
\begin{equation}
 \mathbf{p}_\mathrm{field}=\int \frac{\mathbf{S}(\mathbf{r},t)}{c^2}d^3r=\frac{N_0\hbar\omega}{nc}\hat{\mathbf{x}}.
\end{equation}
The results agree again within the numerical accuracy of the simulations.

The full agreement between the MP and OCD models is only obtained
if the optical force density used in the OCD model is of the Abraham form as given
in Eq.~\eqref{eq:opticalforcedensity}. Therefore, our
results provide extremely strong support for the Abraham force density
as the only optical force density that is fully consistent with the
covariance principle. It is also important to note that
the interface force term in Eq.~\eqref{eq:opticalforcedensity}
is not independent of the Abraham force density in the second term
as these terms are intimately connected by
the conservation law of momentum at interfaces.

\section{\label{sec:discussion}Discussion}

\subsection{\label{sec:significance}Interpretation of the results}

The MP quasiparticle and OCD models
give independent but complementary views of how the covariance principle of
the special theory of relativity governs the propagation of light in a medium.
In the MP picture, the atomic mass transferred
with MDW coupled to the photon becomes quantized and
depends on the energy of the photon as given in Eq.~\eqref{eq:mdwmass1}. In OCD,
the mass is transferred by MDW, which
phenomenologically could also be called an optoelastic shock wave.
The transferred mass per unit energy of the incoming light pulse is,
however, the same in the MP and OCD models.

Our discoveries on the momentum of light are also
fundamental. Both the MP picture and the OCD approach
show that the momentum of a light pulse is carried forward not only by
the field, but also by medium atoms which are set in motion in the
direction of the wave vector by the field-dipole forces.
Actually, for refractive indices $n>\sqrt{2}$,
most of the momentum is carried forward by atoms
as seen in Eq.~\eqref{eq:momentumsplitting} of the MP picture or
in Eqs.~\eqref{eq:dipolemomentum} and \eqref{eq:fieldmomentum} of the OCD approach.
The total momentum of MP becomes equal to the Minkowski momentum.
Therefore, the covariant theory gives a unique and transparent
resolution to the long-standing Abraham-Minkowski controversy.
The difference between the covariant MP and OCD theories and the conventional
theoretical formulations neglecting the optoelastic dynamics of the medium
can be traced back to the conventional initial assumption that
only the field energy propagates in the medium.

Another fundamental
contribution of our work is the analysis of the relaxation phenomena
taking place after the photon transmission. Since our OCD method
includes the elastic forces on the same footing as the field-dipole forces,
we are able to calculate how the mass and thermal equilibria are
gradually reestablished by elastic waves which propagate at the speed
of sound in the medium. When the equilibrium is restored, part of
the field energy has been converted to lattice heat or thermal phonons.
One can and should ask how the photon mass drag effect has remained
undiscovered although the underlying fundamental theories governing it have
been formulated more than a century ago. Any detailed answer deserves
a separate review since previous literature neglecting the dynamics
of the medium under the influence of the optical field is very extensive.
Below we, however, compare our work to
selected earlier theoretical works.

In this work, we have focused on the simulation of Gaussian light pulses
instead of stationary light beams. This is not a
limitation of our theory and the simulation of continuous
laser beams is a topic of further work.
In the case of the simulation of incoherent fields, OCD must be
generalized to account for the field quantization since
the classical fields cannot describe the correlation properties
of chaotic fields.

\subsection{\label{sec:theorycomparison}Comparison with previous theories}

Previous theoretical works have correctly
stated that both the field and the matter parts of the total momentum
are essential in the description of the propagation of light
in a medium \cite{Pfeifer2007,Milonni2005,Barnett2010b,Penfield1967,Gordon1973}.
The separation of momentum into
the field- and matter-related parts presented in some works, e.g., in Eq.~(11) of Ref.~\cite{Milonni2005},
appears to be fully consistent with our
Eqs.~\eqref{eq:dipolemomentum}--\eqref{eq:polaritonmomentum}.
However, none of the previous works have shown
how the momentum taken by the medium gives rise to MDW that propagates
with the light wave in the medium.
Instead, the transfer of mass and the related kinetic and elastic energies
are completely neglected in the previous theories of light propagation in a medium.
Thus, the neglectance of MDW and the related optoelastic dynamics is
the main reason for the long-standing Abraham-Minkowski controversy.
It has, for example, led some researchers to suggest that
the division of the total EMT
into electromagnetic and material components would
be arbitrary as long as the total momentum is uniquely defined \cite{Pfeifer2007}.
This is in contrast with the results of our MP and OCD approaches.
We include into the material part of the total EMT only
physical quantities directly related to the medium like the
velocity distribution of atoms, momentum of atoms,
and the transferred mass. The physical quantities related to the medium are
classical and thus in principle directly measurable.
Therefore, the division of EMT into
field- and material-related parts becomes unambiguous.
To facilitate the comparison with the previous
EMT formulations, we have also presented in Appendix \ref{apx:tensors}
how the results of Sec.~\ref{sec:continuum} can be
formulated into a covariant EMT formalism that fulfills
the conservation laws of energy, momentum, and angular momentum.

In some previous works \cite{Barnett2010b,Barnett2010a},
the Abraham and Minkowski momenta have been related
to the kinetic and canonical momenta of light, respectively.
According to our theory, the MDW momentum in Eq.~\eqref{eq:mdwmomentum}
is the uniquely defined momentum of the MDW atoms, governed by Newton's equation of motion.
Thus, our MDW momentum is both the kinetic and canonical momentum
in the conventional sense \cite{Landau2010}. Also, the field's share of the momentum,
the Abraham momentum, is equal to the conventional kinetic or canonical momentum
of the field \cite{Landau2010}. Therefore, it is impossible to separate
the kinetic and canonical momenta in our theory.
Definition and physical meaning of the canonical momentum of MP
deserves a separate discussion
in the context of a dispersive medium \cite{Barnett2010b,Barnett2010a}.

Some of the most recent works \cite{Leonhardt2014,Zhang2015}
apply fluid dynamics to study the momentum transport of light in fluids.
This approach reminds our OCD theory and appears very promising.
However, in these works,
the dynamics of the fluid is not studied in the time scale of light
propagation as one concentrates on the deformation of the fluid surface due to a
stationary light beam and concludes that neither the Abraham nor the Minkowski
momentum is fundamental, but they emerge depending on the fluid dynamics.
Our work also shows that
the assumption of an incompressible fluid used in these studies fails
if one wants to apply the OCD model to fluids
since the incompressibility makes the medium fully rigid and thus
the force field would propagate at infinite speed, ruining the relativistic
invariance of the theory.

We also want to point out the main differences between our theory and 
previous theories including different photon mass concepts
\cite{Mendonca2000,Zalesny2001,Wang2013a,Wang2013b,Wang2015}.
In previous works,
the concept of the mass of a photon is very abstract as it has not been shown
to be in any way related to the mass density
perturbations in the medium. Particularly, 
the theory by Mendon\ifmmode \mbox{\c{c}}\else \c{c}\fi{}a \emph{et al.} \cite{Mendonca2000}
deals with the special case of plasma and defines the effective mass of the photon
by the dispersion relation of plasma without any resort to the covariance principle.
Therefore, this case is distinctively different from our case of a nondispersive medium.
The neglectance of the transferred
mass $\delta m$ in Zale\ifmmode \acute{s}\else \'{s}\fi{}ny's \cite{Zalesny2001} and Wang's
\cite{Wang2013a,Wang2013b,Wang2015}
theories in turn leads to complicated mathematics
without providing transparent and physically insightful covariant theory of light.
In Zale\ifmmode \acute{s}\else \'{s}\fi{}ny's theory, the
velocity of a photon is neither the
phase velocity nor the group velocity.
The theory by Wang \cite{Wang2013a,Wang2013b,Wang2015}
is especially claimed to be covariant, but it still neglects
the transferred mass $\delta m$, thus leading to mathematical problems, such as
the so-called ``intrinsic Lorentz violation" observed by Wang \cite{Wang2015,Wang2014}.

\subsection{\label{sec:measurements}Experimental verification of the mass transfer}

It is difficult to experimentally quantify the MDW effect by
measuring the recoil effect of atoms at the interfaces.
However, measuring the atomic displacement in the middle of the medium in the time scale
of light propagation would provide a direct proof of the MDW effect.
In this time scale, the recoil momenta taken by the thin interface layers
have not had time to be relaxed by elastic forces. In our one-dimensional example, the constant atomic
displacement inside the medium is 2.7 nm and might be made larger in the optimal experimental setup.
This shift should be within reach using presently available pulse lasers, waveguides,
and micro-optics.
However, in order to get a laser pulse to propagate in a small cross-sectional area,
one would, in reality, need to use an optical waveguide.
In this case, one has to account for the dispersion and the losses
that emerge at the first interface as only a part of the initial
pulse energy can be coupled inside the waveguide.
The waveguide cannot either have a thick coating if the atomic
displacement is measured directly at the surface of the waveguide.
The OCD method can be easily combined with the standard integrated optics
design tools for optimal planning of the experiments.

\section{\label{sec:conclusions}Conclusions}

In conclusion, we have shown that the light pulse propagating in
a nondispersive medium
has to be described as a coupled state of the field and matter.
We have elaborated this coupled state using two different
approaches, the MP quasiparticle picture and the OCD method, which independently prove
the existence of the photon mass drag effect, the transfer of mass with
the light pulse.
To agree with the fundamental conservation
laws of nature and the special theory of relativity,
the light pulse, as a coupled state of the field and matter, must have a finite rest mass and the Minkowski momentum.
The transfer of mass with the light pulse gives rise to nonequilibrium
of the mass density in the medium. When the mass equilibrium is
reestablished by relaxation, a small amount of initial photon
energy is converted to lattice heat. These discoveries change our
understanding of light-matter interaction and our vision of light
in a fundamental way.
We have calculated the mass transfer and dissipation
numerically for one- and three-dimensional Gaussian wave packets
and the diamond crystal with realistic material parameters.
The mass transfer and dissipation are
real-world phenomena that can also be studied experimentally.
We have also shown that an experimental
setup based on the titanium-sapphire pulse laser and 
a waveguide should enable experimental
verification of the mass transfer.
Thus, our work is of great interest to
scientists experimenting with light.

\begin{acknowledgments}
This work has in part been funded by the Academy of Finland
under contract number 287074
and the Aalto Energy Efficiency Research Programme.
\end{acknowledgments}

\appendix

\section{\label{apx:DopplerLorentz}Doppler-Lorentz transformation}

As discussed in Sec.~\ref{sec:mptheory},
in L frame, the energy and momentum of MP have two equivalent expressions, given by
$E_\text{\tiny MP}=\hbar\omega+\delta mc^2=\gamma m_0c^2$ and $p_\text{\tiny MP}=n\hbar\omega/c=\gamma m_0v$.
Therefore, the Lorentz transformation can be written equivalently by using
both expressions. The general form of the Lorentz transformation from L frame
to an arbitrary frame of reference (G frame) moving with velocity $v'$ in L frame
is given in Eqs.~\eqref{eq:LorentzE} and \eqref{eq:Lorentzp}.
Using $E_\text{\tiny MP}=\hbar\omega+\delta mc^2$ and $p_\text{\tiny MP}=n\hbar\omega/c$, we can write
the Lorentz transformation in the form
\begin{align}
 \hbar\omega+\delta mc^2 &\longrightarrow\gamma'\Big(1-\frac{nv'}{c}\Big)\hbar\omega+\gamma'\delta mc^2,\nonumber\\
 \frac{n\hbar\omega}{c} &\longrightarrow\gamma'\Big(1-\frac{nv'}{c}\Big)\frac{n\hbar\omega}{c}.
 \label{eq:DopplerLorentz}
\end{align}
Using $E_\text{\tiny MP}=\gamma m_0c^2$ and $p_\text{\tiny MP}=\gamma m_0v$, we obtain, respectively,
\begin{align}
 \gamma m_0c^2 &\longrightarrow\gamma_\mathrm{rel}m_0c^2,\nonumber\\[10pt]
 \gamma m_0v &\longrightarrow\gamma_\mathrm{rel}m_0v_\mathrm{rel},
 \label{eq:normalLorentz}
\end{align}
where $v_\mathrm{rel}$ is the relative velocity between R frame with velocity $v=c/n$
and G frame with velocity $v'$ and $\gamma_\mathrm{rel}$ is the corresponding
Lorentz factor. The relative velocity $v_\mathrm{rel}$ is obtained by using
the relativistic velocity subtraction as
$v_\mathrm{rel}=(v-v')/(1-vv'/c^2)$. We call the transformation
in Eq.~\eqref{eq:DopplerLorentz} as the Doppler-Lorentz transformation
since the frequency part of the transformation corresponds to the conventional
Doppler shift. The second form of the transformation in Eq.~\eqref{eq:normalLorentz}
is the conventional form of the Lorentz transformation applied
for the moving rest mass $m_0$.

\section{\label{apx:tensors}Energy-momentum tensors}

Here, we present the results of Eqs.~\eqref{eq:dipolemomentum}--\eqref{eq:fieldenergy}
using the well-known EMT formalism.
We split the total EMT of MP into the field and the MDW parts.
The Abraham tensor is found to present the electromagnetic
field part of MP and, when we add to it the
tensor related to MDW, we obtain the total MP tensor
that gives the Minkowski momentum as
the total momentum of MP.

The field momentum and energy in
Eqs.~\eqref{eq:fieldmomentum} and \eqref{eq:fieldenergy}
form the EMT describing the electromagnetic field.
This EMT, which equals the conventional Abraham EMT,
is written as a $4\times 4$ matrix \cite{Pfeifer2007}
\begin{equation}
 \mathbf{T}_\text{field}=
 \bigg[\begin{array}{cc}
  \frac{1}{2}(\varepsilon\mathbf{E}^2+\mu\mathbf{H}^2) & \frac{1}{c}(\mathbf{E}\times\mathbf{H})^T\\
  \frac{1}{c}\mathbf{E}\times\mathbf{H} & -\boldsymbol{\sigma}
 \end{array}\bigg],
 \label{eq:tensorfield}
\end{equation}
where $T$ denotes the transpose and $\boldsymbol{\sigma}$ is
the Maxwell stress tensor given by a $3\times 3$ matrix \cite{Jackson1999}
\begin{equation}
 \boldsymbol{\sigma}=\varepsilon\mathbf{E}\otimes\mathbf{E}+\mu\mathbf{H}\otimes\mathbf{H}
 -\frac{1}{2}(\varepsilon\mathbf{E}^2+\mu\mathbf{H}^2)\mathbf{I}.
\end{equation}
Here, $\otimes$ denotes the outer product and
$\mathbf{I}$ is the $3\times3$ unit matrix.
The EMT in Eq.~\eqref{eq:tensorfield} is traceless which is ralated
to the masslessness of the electromagnetic field \cite{Garg2012}. The Abraham tensor leaves out the energy
and momentum associated to MDW and, thus, they must be incorporated
with a separate MDW tensor. 

Like the Abraham tensor in Eq.~\eqref{eq:tensorfield}, also the MDW tensor,
including the momentum and
energy of MDW in Eqs.~\eqref{eq:dipolemomentum}
and \eqref{eq:dipoleenergy}, must be symmetric as
required by the conservation of angular momentum for the total
EMT of MP.
This tensor, which essentially incorporates the terms
of the total EMT of MP that have been left
out from the Abraham tensor in Eq.~\eqref{eq:tensorfield}, is given by
\begin{align}
 \mathbf{T}_\text{\tiny MDW} &=
 \bigg[\begin{array}{cc}
  \rho_\text{\tiny MDW}c^2 & \rho_\text{\tiny MDW}c\mathbf{v}^T\\
  \rho_\text{\tiny MDW}c\mathbf{v} & \mathbf{0}
 \end{array}\bigg]\nonumber\\
 &=\bigg[\begin{array}{cc}
  \frac{n^2-1}{2}(\varepsilon\mathbf{E}^2+\mu\mathbf{H}^2) & \frac{n^2-1}{c}(\mathbf{E}\times\mathbf{H})^T\\
  \frac{n^2-1}{c}\mathbf{E}\times\mathbf{H} & \mathbf{0}
 \end{array}\bigg],
 \label{eq:tensordipole}
\end{align}
where $\mathbf{0}$ is the $3\times 3$ zero matrix assuming that
the extremely small kinetic energy of MDW is neglected.
In contrast to the Abraham EMT in Eq.~\eqref{eq:tensorfield},
the MDW tensor in Eq.~\eqref{eq:tensordipole} is not traceless due to the term
$\rho_\text{\tiny MDW}c^2=\frac{n^2-1}{2}(\varepsilon\mathbf{E}^2+\mu\mathbf{H}^2)$.
This term integrates to $\delta mc^2$ for a single photon
in analogy with the MP model.

The total EMT of MP is the sum of the
Abraham tensor in Eq.~\eqref{eq:tensorfield}
and the MDW tensor in Eq.~\eqref{eq:tensordipole} and it is given by
\begin{equation}
 \mathbf{T}_\text{\tiny MP}=
 \bigg[\begin{array}{cc}
  \frac{n^2}{2}(\varepsilon\mathbf{E}^2+\mu\mathbf{H}^2) & \frac{n^2}{c}(\mathbf{E}\times\mathbf{H})^T\\
  \frac{n^2}{c}\mathbf{E}\times\mathbf{H} & -\boldsymbol{\sigma}
 \end{array}\bigg].
 \label{eq:tensorpolariton}
\end{equation}
The total EMT of MP in Eq.~\eqref{eq:tensorpolariton}
obeys the conservation laws of both linear and angular momentum such that
$\partial_\alpha T_\text{\tiny MP}^{\alpha\beta}=0$ and $T_\text{\tiny MP}^{\alpha\beta}=T_\text{\tiny MP}^{\beta\alpha}$,
where the greek indices range from 0 to 3 or over $(ct,x,y,z)$ \cite{Jackson1999}.
Verifying that the conservation laws
are satisfied for a field propagating in a medium with constant refractive
index $n$ is straightforward.

In contrast to earlier formulations of the EMTs of the field and matter
\cite{Einstein1908,Chu1966,Fano1968,Sheppard2016,Obukhov2003,Baxter2010,Crenshaw2014,Crenshaw2011},
our total EMT of MP
in Eq.~\eqref{eq:tensorpolariton} includes only the mass density term
related to MDW that is propagating with the light wave.
In addition, in earlier works, one typical starting point is the assumption that
the total momentum of the propagating part of the field
and matter must remain constant
when the light wave enters from vacuum to a medium \cite{Pfeifer2007}.
As we have shown, one can not derive a covariant
energy and momentum from this starting point.
Therefore, detailed comparison of our EMT
of MP and earlier EMTs is not meaningful.

For MP propagating in the $x$ direction,
the total energy is given by
$E_\text{\tiny MP}=\int T_\text{\tiny MP}^{00}d^3r=n^2\hbar\omega=\hbar\omega+\delta mc^2$ and
the magnitude of the MP momentum is given by
$p_\text{\tiny MP}=\frac{1}{c}\int T_\text{\tiny MP}^{10}d^3r=n\hbar\omega/c$.
In the case of the tensors $\mathbf{T}_\text{field}$ and
$\mathbf{T}_\text{\tiny MDW}$
the corresponding quantities are given by
$E_\text{field}=\int T_\text{field}^{00}d^3r=\hbar\omega$,
$E_\text{\tiny MDW}=\int T_\text{\tiny MDW}^{00}d^3r=(n^2-1)\hbar\omega=\delta mc^2$,
$p_\text{field}=\frac{1}{c}\int T_\text{field}^{10}d^3r=\hbar\omega/(nc)$, and
$p_\text{\tiny MDW}=\frac{1}{c}\int T_\text{\tiny MDW}^{10}d^3r=(n-1/n)\hbar\omega/c$,
which are all in accordance with our covariant MP model.

In the description of the photon transmission through a medium
block, we also need the EMT of the recoiling medium block.
This is given by
\begin{equation}
 \mathbf{T}_\mathrm{med}=
 \bigg[\begin{array}{cc}
  \rho_\mathrm{r} c^2 &  \rho_\mathrm{r} c\mathbf{v}_\mathrm{r}^T\\
   \rho_\mathrm{r} c\mathbf{v}_\mathrm{r} & \rho_\mathrm{r}\mathbf{v}_\mathrm{r}\otimes\mathbf{v}_\mathrm{r}
 \end{array}\bigg],
\end{equation}
where $\rho_\mathrm{r}=\rho_0+\rho_\mathrm{rec}$ is the mass density of the
recoiling medium and $\mathbf{v}_\mathrm{r}$ is the very small
recoil velocity field.
The corresponding energy and momentum are given by
$E_\mathrm{med}=\int T_\mathrm{med}^{00}d^3r=M_\mathrm{r}c^2$ and
$p_\mathrm{med}=\frac{1}{c}\int T_\mathrm{med}^{10}d^3r=M_\mathrm{r}V_\mathrm{r}$,
where $M_\mathrm{r}=\int\rho_\mathrm{r}d^3r$ and
$V_\mathrm{r}=\int\rho_\mathrm{r}v_\mathrm{r}d^3r/M_\mathrm{r}$.

\vspace{0.4cm}
\section{\label{apx:flowchart}Computational details}

In our simulations, the fields are calculated from the
analytic solutions described in Secs.~\ref{sec:1dsimulations} and
\ref{sec:3dsimulations} piecewise in vacuum and in diamond. 
The effects of the Gaussian pulse on the material are in turn calculated
numerically using Newton's equation of motion in Eq.~\eqref{eq:mediumnewton} and
the optical and elastic force densities in Eqs.~\eqref{eq:opticalforcedensity}
and \eqref{eq:elasticforcedensity}.
Adopting this perturbative approach is justified
as the effects of the fields on the state of the material are
extremely small, and the back action of the changes in
the state of the material on the field is even smaller (see Sec.~\ref{sec:3dsimulations}).

For the simulations, we first choose a time step $h_t$ that is
chosen to be small compared to the temporal width of the pulse
or the harmonic cycle depending on the calculation. In the one-dimensional
calculations, we use the discretization length $h_x$ with respect to position,
and in the three-dimensional case, we discretize the medium to voxels
with side lengths denoted by $h_x$, $h_y$, and $h_z$.
Depending on the calculation, these are chosen to be small compared
to the dimensions of the pulse in space or the scale
of the harmonic cycle.
We tested the influence of the grid size on the computed MDW mass
and momentum and chose the grid size so that it gives a
relative numerical error smaller than $10^{-7}$.
We were also able to reproduce the results of the MP quasiparticle
model within this numerical accuracy.
The volume elements of the one- and three-dimensional
calculations are described in more detail below.
We start the simulation 100 ps before the center of the Gaussian light pulse
has reached the position of the first interface of the simulation
geometry in Fig.~\ref{fig:simulationgeometry}.
The iteration loop in the simulation consists of the following steps:
\vspace{0.2cm}
\begin{enumerate}
 \item Using Eqs.~\eqref{eq:opticalforcedensity} and \eqref{eq:elasticforcedensity},
       compute the optical and elastic forces experienced by the medium
       elements with a one-dimensional width $h_x$ or three-dimensional volume $h_xh_yh_z$.
\vspace{0.2cm}
 \item Using time step $h_t$, calculate the acceleration, velocities, and positions
       of the medium elements according to the forces and
       Newton's equation of motion in Eq.~\eqref{eq:mediumnewton}. Return to step 1.
\end{enumerate}
After the light pulse has left the medium, we increase the time step to
$h_t=1$ ns, which is still small compared to the time scale of the elastic forces. Then, we
continue the simulation to see the relaxation of the mass nonequilibrium
resulting from the MDW and recoil effects.

In the present simulations, we have described the refractive index
as a step function near the surfaces thus neglecting any atomic
scale changes in the refractive index.
If one assumes perfect transmission, in this approximation,
the integral of the surface force density, described by
the first term in the optical force density in Eq.~\eqref{eq:opticalforcedensity},
becomes $-\int\frac{\varepsilon_0}{2}E^2\nabla n^2d^3r\approx-(1/n_\mathrm{L}-1/n_\mathrm{R})\int u_\mathrm{R}dydz$.
Here $n_\mathrm{L}$ and $n_\mathrm{R}$ are the refractive indices on
the left and right of the surface and $u_\mathrm{R}$ is the electromagnetic
energy density on the right of the surface. In the present simulations,
it is again averaged over the harmonic cycle.
Furthermore, the grid
used in the simulations is so coarse that the whole recoil energy
and momentum is taken into volume of the first grid layer.
Since the recoil energy depends on the recoiling mass,
this means that the recoil energy obtained from the simulations
is only approximative. However, independently on the grid size,
the recoil energy remains negligibly small and thus does not have any relevance
regarding the total energy of the light pulse. The calculations also show
that the recoil momentum does not depend on the grid size.
As seen in Figs.~\ref{fig:mdw} and \ref{fig:displacementsimulation},
the peaks associated with the recoil displacement
of atoms at the interfaces are not accurate but only qualitative.
This is, however, not a fundamental limitation of OCD, but a choice made to speed up the simulations.
By increasing computing time and including the surface effects
on the refractive index, the calculations could be repeated to
an accuracy that is sufficient to describe the recoil effect quantitatively
within the limits of the classical OCD theory.

\end{document}